\documentclass[iop]{emulateapj}
\usepackage{graphicx}
\usepackage{amssymb}
\usepackage[T1]{fontenc}
\usepackage{aecompl}

\shorttitle{The Deuterium Fractionation Timescale}
\shortauthors{Kong et al.}

\begin{document}

\title{The Deuterium Fractionation Timescale in Dense Cloud Cores: A Parameter Space Exploration}

\author{Shuo Kong\altaffilmark{1}}
\affil{Dept. of Astronomy, University of Florida, Gainesville, Florida 32611, USA}
\email{skong@astro.ufl.edu}

\author{Paola Caselli\altaffilmark{2,3}}
\affil{Max-Planck-Institute for Extraterrestrial Physics (MPE), Giessenbachstr. 1, D-85748 Garching, Germany}
\affil{School of Physics and Astronomy, University of Leeds, Leeds LS2 9JT, UK}
\email{caselli@mpe.mpg.de}

\author{Jonathan C. Tan\altaffilmark{1,4}}
\affil{Dept. of Astronomy, University of Florida, Gainesville, Florida 32611, USA}
\affil{Dept. of Physics, University of Florida, Gainesville, Florida 32611, USA}
\email{jt@astro.ufl.edu}

\author{Valentine Wakelam\altaffilmark{5,6}}
\affil{University of Bordeaux, LAB, UMR 5804, 33270, Floirac, France}
\affil{CNRS, LAB, UMR 5804, 33270, Floirac, France}
\email{wakelam@obs.u-bordeaux1.fr}

\and

\author{Olli Sipil\"a\altaffilmark{2}}
\affil{Max-Planck-Institute for Extraterrestrial Physics (MPE), Giessenbachstr. 1, D-85748 Garching, Germany}
\email{olli.sipila@helsinki.fi}


\begin{abstract}

The deuterium fraction [N$_2$D$^+$]/[N$_2$H$^+$], may provide information
about the ages of dense, cold gas structures, important to compare
with dynamical models of cloud core formation and evolution. Here we
introduce a complete chemical network with species containing up to
three atoms, with the exception of the Oxygen chemistry, where reactions 
involving H$_3$O$^+$ and its deuterated forms have been added, 
significantly improving the consistency with comprehensive chemical networks. 
Deuterium chemistry and spin states of H$_2$ and H$_3^+$ isotopologues
are included in this primarily gas-phase chemical model. We
investigate dependence of deuterium chemistry on model parameters:
density ($n_{\rm H}$), temperature, cosmic ray ionization rate, and
gas-phase depletion factor of heavy elements ($f_{\rm D}$).
We also explore the effects of time-dependent freeze-out of 
gas-phase species and dynamical evolution of density at various
rates relative to free-fall collapse. 
For a broad range of model parameters, the timescales to
reach large values of $D_{\rm frac}^{\rm N_2H^+} \ga 0.1$,
observed in some low- and high-mass starless cores,
are relatively long compared to the local free-fall timescale. 
These conclusions are unaffected by introducing time-dependent freeze-out 
and considering models with evolving density, unless the initial $f_{\rm D}
\ga$ 10. For fiducial model parameters, achieving $D_{\rm frac}^{\rm
N_2H^+} \ga 0.1$
requires collapse to be proceeding at rates at least several times slower than
that of free-fall collapse, perhaps indicating a dynamically important
role for magnetic fields in the support of starless cores and thus the
regulation of star formation.
\end{abstract}


\section{Introduction} 

Deuterated molecules are useful diagnostic tools to study the cold and
dense environments where stars are born. This has been demonstrated in
low-mass star-forming regions
\citep[e.g.,][]{Caselli2002,Bacmann2003,Crapsi2005,Crapsi2007,Emprechtinger2009,Friesen2010},
as well as in regions thought to be precursors of massive stars and
stellar clusters
\citep[e.g.,][]{Fontani2006,Fontani2009,Fontani2011,2007A&A...467..207P,2012ApJ...751..135P}. Deuterated
species can be used to infer the elusive electron fraction $x(e)$
\citep[e.g.,][although the equations in these papers need to be
  modified to include the doubly and triply deuterated forms of
  H$_3^+$]{1977ApJ...217L.165G,1979ApJ...234..876W,Caselli1998,Bergin1999,Dalgarno2006}
and the age of molecular clouds \citep{2011ApJ...739L..35P,2013A&A...551A..38P,2014Natur.516..219B}. 
Electron fraction and cloud age are two
important parameters to shed light on the dynamical evolution of
star-forming regions, as the ambipolar diffusion timescale is directly
proportional to $x(e)$ \citep[e.g.,][]{1987ARA&A..25...23S} and the age can put
stringent constraints on the mechanism(s) regulating cloud core
formation (e.g., magnetic fields, turbulence and shocks).  However,
variations in cosmic-ray (CR) ionization rate, volume density, kinetic
temperature, rates of molecular freeze-out onto dust grain surfaces
and the ortho-to-para ratio of H$_2$ make attempts to fix these values
rather uncertain, especially for regions with poorly known physical
structure.

Extensive effort have been spent on understanding the chemistry
in starless/pre-stellar cores
(e.g. \citealt{Flower2006}; \citealt{2009A&A...494..623P}, hereafter P09;
\citealt{2011A&A...526A..31P}; \citealt{aikawa2012}; \citealt{2013A&A...554A..92S}).
However, since they focused on specific aspects of modeling, 
they were limited by either the incompleteness of reactions or
the narrow range of physical conditions. In this paper we use a
complete reduced network with up-to-date rate coefficients, 
and explore uniformly the parameter space without any
prior bias about the dynamical history, as this depends on poorly known 
physical quantities such as magnetic fields and turbulence. 
This parameter space exploration is needed to understand the dependence 
of the chemical composition (in particular the abundance of deuterated molecules) 
on basic physical properties and parameters, and to help the interpretation 
of observational data.  It is the first time such an exploration
has been done with complete spin-state reactions.

Recently, \citet{2013A&A...551A..38P} investigated these effects by coupling
hydrodynamics with chemistry.  They developed an astrochemical model
to derive the age of low-mass cores and extensively discussed the role
of o-H$_2$. However, they used a
relatively limited set of reactions: their chemical network was first
based on that of \citet{2005A&A...443..961L} with 120 reactions and 35 species;
then later improved by P09 to include
about 400 reactions. The P09 network ignores reactions with rate
coefficients below 10$^{-15}$ cm$^{-3}$ s$^{-1}$, thus no radiative association 
reactions, important for carbon chemistry, are included.
Furthermore, their model does not fully track the N chemistry ($\rm
N_2$ abundance is a parameter), so they are not able to predict
absolute abundances of N$_2$H$^+$ and N$_2$D$^+$.  

In this paper, we first introduce a more complete chemical network and
describe our methods of following chemical evolution (\S\ref{descrip}).
Then in \S\ref{sec:results} we present our results for determining the
chemical age of cloud cores by the deuterium fraction of
N$_2$H$^+$. This is similar to the approaches of P09 and
\citet{2013A&A...551A..38P}, but extended to cover a broader range
of conditions, including those relevant to high-mass star-forming
regions that may contain massive starless cores (e.g., Tan et
al. 2013). Furthermore, we consider a range of simple parameterized
collapse rates relative to that of free-fall collapse.
The implications of our results are discussed in
\S\ref{sec:discussion}, including detailed comparison with the results
of P09 and \citet{2013A&A...551A..38P}.  Conclusions are summarized in
\S\ref{sec:conclusion}.

\section[]{Methods} \label{descrip} 

\subsection{Fiducial Chemical Network}

Our model is based on the network first described in \citet{2012A&A...547A..33V} 
(hereafter V12), who used the chemistry results to 
interpret ground-based and Herschel Space Observatory observations of 
deuterated isotopologues of H$_3^+$ toward a pre-stellar core 
(see description in their \S3.3).  The V12 code was originally built starting from 
a complete reaction network including only molecules with up to three atoms in size, 
extracted from the Nahoon network \citep{2012ApJS..199...21W}, which is available in 
the KIDA\footnote{http://kida.obs.u-bordeaux1.fr/} database (Oct. 2010 version).  
The reduced network only includes the elements H, D, He, O, C and N. 
This simplified network still allows us to follow easy-to-observe species in the gas phase, 
such as N$_2$H$^+$, HCO$^+$ and their deuterated forms. 
The reduced network includes the spin states of H$_2$, H$_3^+$ and their 
deuterated isotopologues, following prescriptions of
\citet{2004A&A...418.1035W}, \citet{Flower2006}, \citet{Hugo2009}, 
P09, \citet{2010A&A...509A..98S}, and
selecting the most recent values for the rate coefficients from the 2010 KIDA database. 

We have made five main improvements to the V12 network: 
(1) The dissociative recombination rates of all the forms of
H$_3^+$ have been calculated through the interpolation of 
Table B.1 of P09.   
(2) Rate coefficients have been updated and recombination reactions of 
C$^+$, N$^+$ and O$^+$ onto negatively charged dust grains have been added, 
following the more recent 2011 KIDA network.  
(3) Bugs in the duplication routine used to construct the V12 network have been corrected. 
In particular, we adjusted the branching ratio of reactions such as 
\begin{equation}\label{equ:branching1}
\rm{HD~+~CO^+~\rightarrow~H~+~DCO^+}
\end{equation}
\begin{equation}\label{equ:branching2}
\rm{HD~+~CO^+~\rightarrow~D~+~HCO^+}, 
\end{equation}
which is now a half of that of the following reaction from KIDA:
\begin{equation}\label{equ:nobranching}
\rm{H_2~+~CO^+~\rightarrow~H~+~HCO^+}.
\end{equation}
(4) We checked our network against that of \citet{2013A&A...554A..92S} (hereafter S13)
to make sure that spin-state rules were followed. 
This implied the elimination of some reactions, such as charge exchange reactions 
involving spin changes; the elimination of reactions of the type 
\begin{equation}\label{equ:forbidden1}
\textrm{H}_2\textrm{O}~+~\textrm{H}^-~\rightarrow~\textrm{o-H}_2~+~\textrm{OH}^-, 
\end{equation}
as in cold gas it is assumed that only p-H$_2$ can form in reactions 
containing only reactants other than H$_2^+$ , H$_2$ and H$_3^+$; 
the elimination of a few reactions built by the V12 duplication code 
which did not follow \citet{2004JMoSp.228..635O} spin rules, such as 
\begin{equation}\label{equ:forbidden2}
\textrm{p-H}_2~+~\textrm{p-D}_2^+~\rightarrow~\textrm{H}~+~\textrm{o-D}_2\textrm{H}^+.   
\end{equation}
(5) Because of its importance for Oxygen chemistry, we include 
$\rm H_3O^+$ and its deuterated isotopologues, as well as all reactions 
involving species that are present in our network.  The inclusion of $\rm H_3O^+$ 
significantly improves the overall agreement with S13.
The abundances of electrons, water, 
CO, $\rm HCO^+$, $\rm DCO^+$, $\rm N_2$, $\rm N_2H^+$, $\rm N_2D^+$
(the most important species in the network, as the deuterium fraction is typically 
measured through the N$_2$D$^+$/N$_2$H$^+$ 
and/or DCO$^+$/HCO$^+$ column density ratios)
are always within a factor of 2 when compared to S13 network. 
This is also true for deuterium fraction
and its equilibrium timescale (defined and studied later in \S\ref{subsec:dfrac}).
We did not include the surface chemistry described in S13,
since there are large uncertainties involved, 
while not significantly impacting the gas-phase chemistry of 
H$_3^+$, HCO$^+$, N$_2$H$^+$ and their deuterated forms 
in cold regions (see \S\ref{S:TDD}). 
However, the surface formation of p-H$_2$, o-H$_2$, HD,
p-D$_2$ and o-D$_2$ are included in our network. 
The rates have been calculated following \citet{2002A&A...390..369L}.
The ortho-to-para ratio upon surface formation
has been assumed equal to the statistical value of 3 for H$_2$ and 2
for D$_2$.  Neutral and negatively charged grains are considered.
Coulomb focusing was taken into account for reactions involving
positively charged ions on negatively charged grains
\citep{Draine1987}. 
Our fiducial chemical network now includes 3232 reactions 
involving 132 different species. 
The network will be publicly available via the KIDA database.  

Our first treatment of molecular freeze-out involves an approximation
of reducing the initial elemental abundances of species heavier than
He by a ``depletion factor'', $f_{\rm D}$ (= 10 for the fiducial model,
fixed in each run). Below, we also describe an extension of this
simple approximation to include time-dependent depletion and
desorption (\S\ref{subsec:realdeplet}).

\subsection[]{Time-Dependent Depletion/Desorption} \label{subsec:realdeplet}

The inclusion of time-dependent depletion and desorption rates of the
heavier elements adds
additional uncertainty and complexity to the modeling (which is why in
the fiducial network, above, depletion factor is treated as a
controllable parameter).
However, in order to gain a basic insight into the potential effects
of these more complex processes, we developed a second network that
includes freeze-out and desorption of neutral species, following
\citet{1992ApJS...82..167H} and \citet{1993MNRAS.261...83H}. 
Hereafter we refer
to this as the Time-Dependent Depletion (TDD) network. Three types of
reactions are implemented: (a) sticking onto dust grains; (b) thermal
evaporation; (c) CR induced evaporation. Binding energies and
sticking coefficients are the same as those used in
\citet{2007A&A...467.1103G}. Altogether there are 153 new reactions added into
the TDD network.

\subsection[]{Fiducial Initial Conditions and Model Parameters} \label{subsec:fiducialmodel}

We choose the following fiducial initial conditions and model
parameters (see also Table \ref{tab:ipc}).
The density is expressed via the number density of H
nuclei, $n_{\rm H}=10^5\:{\rm cm}^{-3}$, gas temperature, $T=15$~K and
heavy element depletion factor, $f_{\rm D}$=10.
The choices of these fiducial values are motivated by observations of
both low- and high-mass pre-stellar cores
\citep[e.g.][]{1999MNRAS.305..143W,Crapsi2005,Crapsi2007,2006A&A...450..569P,2011ApJ...738...11H,2011ApJ...736..163R,BT2012}.
The CR ionization rate, $\zeta=2.5\times 10^{-17}\:{\rm
  s}^{-1}$, is adopted from \citet{2000A&A...358L..79V}.

The fiducial visual extinction, $A_{V}$, is set to 30 mag, a value
large enough so that photochemistry is unimportant for our adopted
radiation field (standard Habing field, $G_0=1$).  
We assume that refractory metals of low ionization
potential (such as Mg and Fe) and polycyclic aromatic hydrocarbons
(PAHs), important for the ionization structure, are not present in
gas phase because of freeze-out onto dust grains, a reasonable assumption 
in cold molecular clouds \citep[see][for the effects of metals and PAHs,
respectively, on the chemical structure of molecular
clouds]{Caselli1998,2008ApJ...680..371W}. The dust-to-gas mass ratio, grain
radius and grain density are taken from the original Nahoon model and
represent the fiducial values typically adopted in chemical models.

\begin{deluxetable}{ccc}
\tabletypesize{\scriptsize}
\tablecaption{Fiducial Parameter Values.\label{tab:ipc}}
\tablewidth{0pt}
\tablehead{
\colhead{Parameter} & \colhead{Description} & \colhead{Fiducial value}
}
\startdata
$n_{\rm H}$ & number density of H nuclei & 1.0$\times$10$^5$ cm$^{-3}$\\
$T$ & temperature & 15 K\\
$\zeta$ & CR ionization rate & 2.5$\times$10$^{-17}$ s$^{-1}$\\
$f_{\rm D}$ &	depletion factor & 10\\
$G_0$ & ratio to Habing field & 1\\
$A_V$ &	visual extinction & 30 mag\\
DGR\tablenotemark{a} & dust-to-gas mass ratio & 7.09$\times10^{-3}$\\
$a_0$ & dust particle radius & 1.0$\times$10$^{-5}$ cm\\
$\rho_{\rm GRAIN}$ & dust grain density & 3.0 g cm$^{-3}$\\
\enddata
\tablenotetext{a}{Following \citet{2011piim.book.....D}.}
\end{deluxetable}

The fiducial initial fractional abundances of elements, with respect
to total H nuclei are listed in Table \ref{tab:ia}. For simplicity,
all species are assumed to be in atomic form, except for H and D.
Deuterium is initially assumed to be all in HD, with a fractional
abundance adopted from the measurement of the elemental [D]/[H] ratio
measured in the Galactic interstellar medium \citep[{[D]/[H] $\sim$
    1.5$\times$10$^{-5}$; e.g.}][]{2003ApJ...587..235O}.
Below, we also investigate the effects of changing these initial
chemical states, finding that our main results are quite insensitive
to these choices.

The fiducial initial ortho-to-para H$_2$ ratio, OPR$^{\rm H_2}$, is
set to its statistical value of 3, assumed to be obtained in the
process of H$_2$ formation on dust grain surfaces. This choice does
impact deuterium chemistry, and so below we do consider the effects of
a range of initial values.

\begin{deluxetable}{cc}
\tabletypesize{\scriptsize}
\tablecaption{Fiducial initial elemental abundances.\label{tab:ia}}
\tablewidth{0pt}
\tablehead{
\colhead{Species} & \colhead{Abundance ( $n_{\rm species}$/$n_{\rm H}$)}
}
\startdata
p-H$_2$ & 1.25$\times$10$^{-01}$ \\
o-H$_2$ & 3.75$\times$10$^{-01}$ \\
HD & 1.50$\times$10$^{-05}$ \\
He & 1.00$\times$10$^{-01}$ \\
N & 2.10$\times$10$^{-06}$ \\
O & 1.80$\times$10$^{-05}$ \\
C & 7.30$\times$10$^{-06}$ \\
GRAIN0 & 1.32$\times$10$^{-12}$ 
\enddata
\end{deluxetable}

\section{Results} \label{sec:results}

\subsection{The Fiducial Model}\label{subsec:fiducial}

\begin{figure*}
\epsscale{1.18}
\plotone{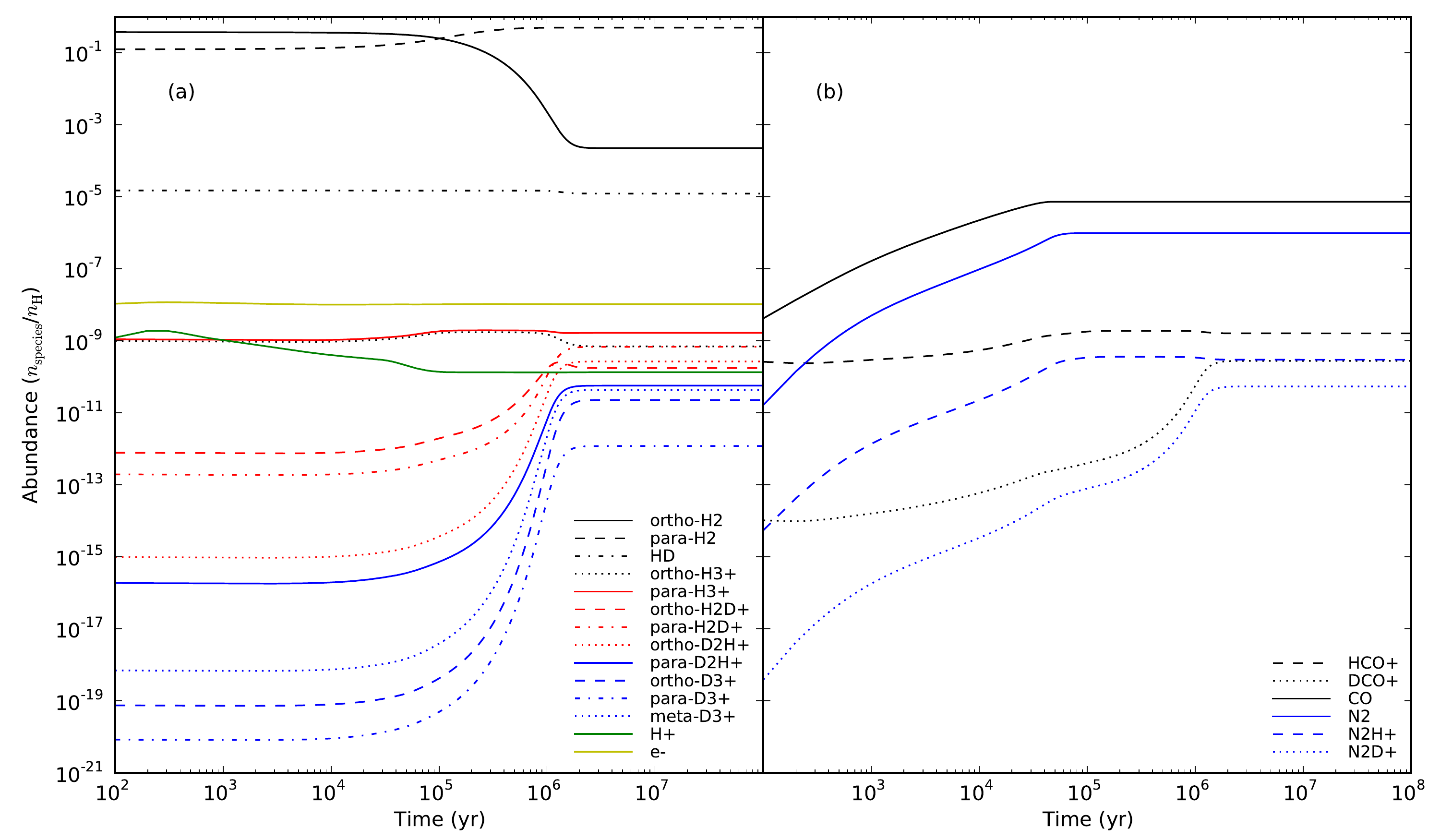}
\caption{%
Time evolution of fractional abundances of important species in the
fiducial model with $n_{\rm H}=$1.0$\times$10$^5$ cm$^{-3}$, $T=$15~K,
$\zeta=$2.5$\times$10$^{-17}$ s$^{-1}$, $f_{\rm D}$=10, $G_0=1$ and $A_V=30$
mag. \textbf{(a):} hydrogen species, including H$_2$, H$_3^+$ and
their deuterated isotopologues (plus spin states), and
electrons. \textbf{(b):} Species of our interest, 
especially N$_2$D$^+$, N$_2$H$^+$, and their progenitor N$_2$.\label{fig:fm}}
\end{figure*}

Figure~\ref{fig:fm} shows the fractional abundances ([species]=
$n_{\rm species}$/$n_{\rm H}$) of important species as a function of time in
the fiducial model, i.e. the fiducial network with fiducial initial
conditions. As the gas
evolves under these cold, dense conditions, the deuteration becomes
active through the exothermic reaction \citep[only true with respect
  to para states of reactants and products;][]{1992A&A...258..479P}:
\begin{equation}\label{equ:d}
\textrm{p-H}_3^+ + \textrm{HD} \rightleftharpoons \textrm{p-H}_2\textrm{D}^+ + \textrm{p-H}_2 + 232\:{\rm K}.
\end{equation}

H$_2$D$^+$ can cede a deuteron to major neutral species, such as CO
and N$_2$, producing DCO$^+$ and N$_2$D$^+$, respectively. As a
consequence, the deuterium fraction (i.e., defined by the abundance
ratios [N$_2$D$^+$]/[N$_2$H$^+$], [DCO$^+$]/[HCO$^+$]) starts to
overcome the cosmic abundance of deuterium. Hereafter, we denote the
deuterium fraction of a certain species as $D_{\rm frac}^{\rm
  species}$ (e.g. [N$_2$D$^+$]/[N$_2$H$^+$] $\equiv$ $D_{\rm
  frac}^{\rm N_2H^+}$) and the spin-state ratio as OPR$^{\rm species}$
(e.g., [o-H$_2$]/[p-H$_2$] $\equiv$ OPR$^{\rm H_2}$).  We will
focus on $D_{\rm frac}^{\rm N_2H^+}$ in our study, since HCO$^+$
suffers more from depletion than N$_2$H$^+$, 
so that $D_{\rm frac}^{\rm N_2H^+}$ is a better tool for tracing the
inner, denser regions of starless/pre-stellar cores
\citep{Casellietal2002,Crapsi2005}.

\begin{figure}
\epsscale{1.15}
\plotone{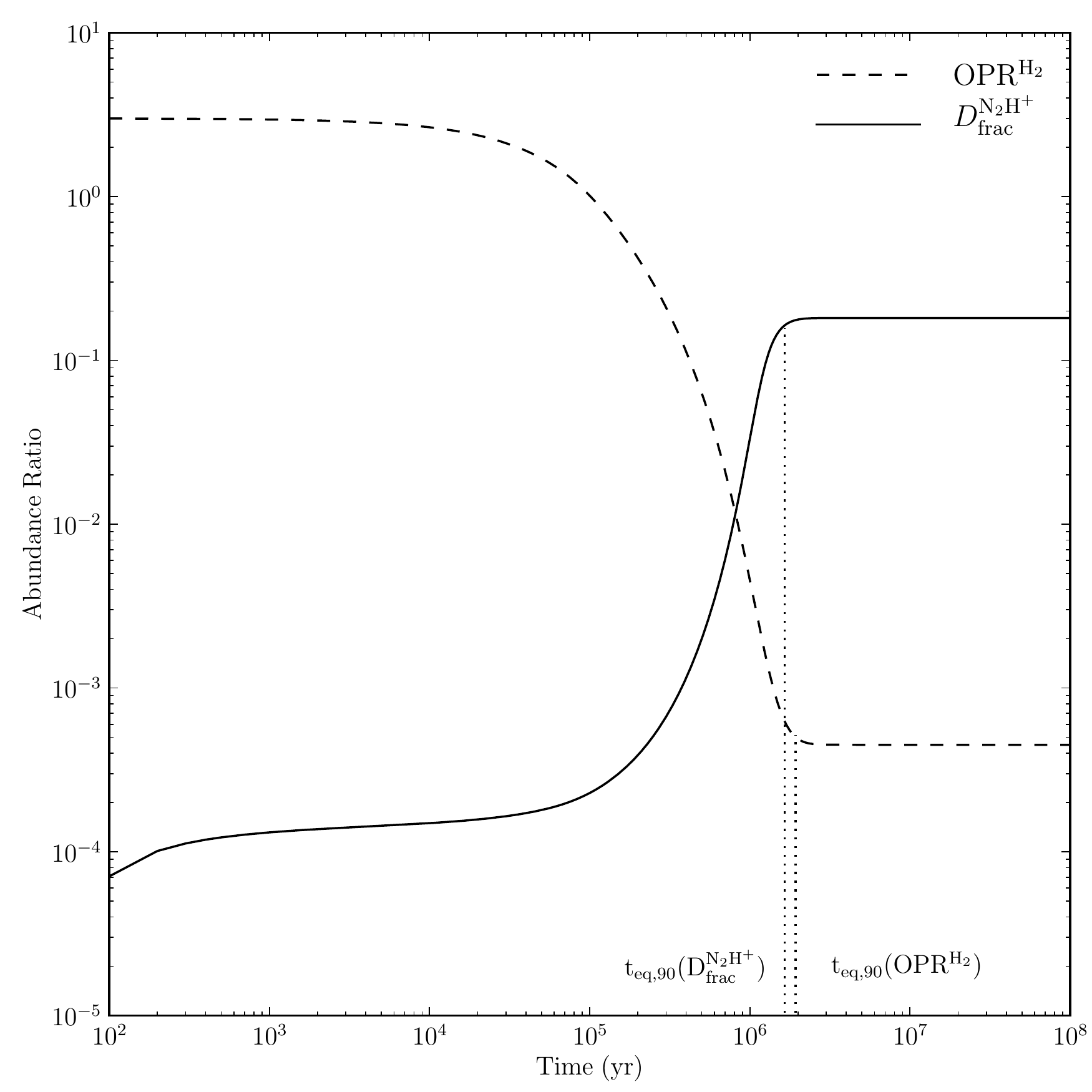}
\caption{%
Time evolution of OPR$^{\rm H_2} \equiv$ [o-H$_2$]/[p-H$_2$] 
and $D_{\rm frac}^{\rm N_2H^+}\equiv$ [N$_2$D$^+$]/[N$_2$H$^+$] in the fiducial model. 
Two dotted lines mark the times when these quantities approach within
10\% of their final equilibrium values ($t_{\rm eq,90}$).\label{fig:fd}}
\end{figure}

The deuterium fraction, shown in Figure~\ref{fig:fd}, increases
significantly only at times later than $\sim$10$^5$\,yr, when the
abundance of o-H$_2$ starts to drop. Deuteration is suppressed by
o-H$_2$, which drives the reaction (\ref{equ:d}) backwards, as
originally pointed out by \citet{1991MNRAS.248..173P} (for p-H$_2$D$^+$) and
\citet{1992A&A...258..479P} (for o-H$_2$D$^+$), and later discussed by
\citet{Flower2006}, P09, and \citet{2011ApJ...739L..35P}.  The conversion of
o-H$_2$ to p-H$_2$ mainly proceeds through the reactions of
o-H$_2$ with H$^+$ and H$_3^+$.  Figure~\ref{fig:fd} shows how
OPR$^{\rm H_2}$ and $D_{\rm frac}^{\rm N_2H^+}$ change together in the
fiducial model: as expected, $D_{\rm frac}^{\rm N_2H^+}$ goes up as
OPR$^{\rm H_2}$ drops. After reaching the equilibrium steady-state at
about $2$ million years, $D_{\rm frac}^{\rm N_2H^+}$ has increased by
about 4 orders of magnitude relative to the cosmic deuterium to
hydrogen abundance ratio, while OPR$^{\rm H_2}$ has dropped by more
than 3 orders of magnitude. One can also see from Figure~\ref{fig:fm}
that all species reach steady-state when OPR$^{\rm H_2}$ does. These
results emphasize that OPR$^{\rm H_2}$ is crucial for cold gas
chemistry in general, for deuterium fractionation in particular and
for the chemical timescale (see \S\ref{subsubsec:iop}).

One thing to note is that in our model the abundance of heavy molecules
such as CO and N$_2$ increase with time
(Fig.\,\ref{fig:fm}). As CO and N$_2$ are both important destruction
partners of H$_3^+$ and its deuterated isotopologues, their increasing
abundance would tend to reduce that of these species. However, the
countervailing effect of the decreasing abundance of o-H$_2$ is
more dominant.
At the physical conditions of the fiducial model, as time proceeds,
species like CO should suffer from increasing amounts of freeze-out
onto dust grains \citep[e.g.][]{Caselli1999}. We have not included any
differential freeze-out mechanism for CO and N$_2$, as laboratory work
has found similar sticking coefficients and binding energies for the
two molecules \citep{Bisschop2006}. Recall also that this fiducial
network assumes a fixed, heavy element-independent depletion factor is
present from the initial condition. The effects of relaxing this
assumption are investigated in \S\ref{S:TDD}.

With these caveats in mind, we note from Figure~\ref{fig:fm}, that the
N$_2$H$^+$/CO ratio increases with time, up to a few times 10$^5$\,yr,
when the N$_2$H$^+$ abundance reaches steady state, as N$_2$, the
precursor molecule to N$_2$H$^+$, forms more slowly than CO, via
neutral-neutral reactions rather than ion-neutral reactions \citep[see
  also][]{2010A&A...513A..41H}.

\subsection{The Deuteration Timescale}\label{subsec:dfrac}

Studies have suggested a theoretical relation between the deuterium
fraction and the evolutionary stage in low-mass cores
\citep{Caselli2002,Crapsi2005,2013A&A...551A..38P}, with the level of
deuteration rising with increasing age and density of the starless
core, before then falling once a protostar forms and starts to heat
its natal envelope.  \citet{Fontani2011} have examined a similar
relation in massive cores, and their findings support the use of
deuterium fraction as an evolutionary indicator for massive starless
and star-forming cores.  

Here we investigate the absolute timescale for the growth of the
deuterium fraction and its implication for the ages of low-mass and
massive starless cores. We also examine how the variation of physical
properties of the gas, including choices of initial conditions,
influences this deuteration timescale, i.e. a chemical timescale. 
For convenience, when considering the output of our chemical network,
we define the equilibrium deuterium fraction, $D_{\rm frac,eq}$ as
the average of two adjacent outputs of $D_{\rm frac}$ (separated by
$\Delta t = 10^4\:{\rm yr}$) that have a fractional change 
\begin{equation}\label{eq:equi}
|\Delta D_{\rm frac}| / D_{\rm frac} < \epsilon,
\end{equation}
with a choice of $\epsilon = 5\times 10^{-5}$.  In practice, we run
the model for 10$^8$ yr and then search backwards in time for when
this condition is satisfied.  We denote the timescale to reach the
equilibrium condition defined by Eq.~(\ref{eq:equi}) as $t_{\rm
  eq}$($D_{\rm frac}^{\rm species}$).  The equilibrium value of the
ortho-to-para ratio of H$_2$, OPR$_{\rm eq}^{\rm H_2}$, is defined in
a similar way, and the timescale is denoted $t_{\rm eq}$(OPR$^{\rm
  H_2}$).  In practice, since the evolution of $D_{\rm frac}^{\rm
  N_2H^+}$ and OPR$_{\rm eq}^{\rm H_2}$ are very slow as they approach
equilibrium (e.g. Fig. \ref{fig:fd}), we also define a more
representative equilibrium timescale $t_{\rm eq,90}$($D_{\rm
  frac}^{\rm N_2H^+}$) as the time when $D_{\rm frac}^{\rm N_2H^+}$
increases to 90\% of $D_{\rm frac,eq}^{\rm N_2H^+}$. In a similar way,
we define $t_{\rm eq,90}$(OPR$^{\rm H_2}$) as the time when OPR$^{\rm
  H_2}$ decreases to OPR$_{\rm eq}^{\rm H_2}$/0.90.

We will compare these chemical timescales to physical timescales, in
particular the local free-fall timescale, $t_{\rm ff}$, which, 
for a uniform density core, is
\begin{equation}\label{eq:ff}
t_{\rm ff} = \left(\frac{3\pi}{32G\rho}\right)^{1/2} = 1.39\times10^5~\left(\frac{n_{\rm H}}{10^5~{\rm cm}^{-3}}\right)^{-1/2}~{\rm yr}. 
\end{equation}
Note that this timescale is evaluated with reference to the current
density of a core, predicting how long it will take in the future to
collapse to a very high density state in the absence of any internal
pressure support. 
However, this timescale is also an approximate
estimate for the minimum amount of time that the core has existed at
densities similar to its current value, since if contraction is driven
by self-gravity we do not expect evolution in core properties to be
proceeding on timescales shorter than the local free-fall
time. 

Furthermore, depending on the degree of turbulent and magnetic
field support, the contraction could be proceeding at rates much
slower than that of free-fall collapse.
Thus, in \S\ref{S:DDE}, we will also consider models in which the density
evolves continuously at various rates relative to the local free-fall time.

Note that the deuteration timescale refers to the age
of a core, which in our fiducial modelling is the time spent at the
given constant density. In the evolving density models, the deuteration timescale
is the time the core has spent evolving from a particular lower
density initial condition to the current density. When comparing to observations, 
one has to take into account that dense cores have been evolving from lower
densities, so models with evolving density structures are important 
to constrain the rates of collapse from the measured abundances of deuterated molecules.

The first line of Table \ref{tab:dfrac} lists equilibrium abundance
ratios and timescales for the fiducial model, i.e. with $n_{\rm H}$ =
10$^5$ cm$^{-3}$. The deuteration timescale $t_{\rm eq,90}$($D_{\rm
  frac}^{\rm N_2H^+}$) is $\simeq$12~$t_{\rm ff}$. Thus, if a starless
core were to be observed with physical and environmental properties
equal to the fiducial model, and $D_{\rm frac} \gtrsim 0.1$, then our
modeling implies it would need to be substantially older than its
current local $t_{\rm ff}$, assuming it had started with our adopted
initial conditions, including the initial OPR of $\rm H_2$. 

\subsection{Effect of Initial Conditions on the Deuteration Timescale}\label{subsec:ic}

\subsubsection{Initial Elemental Abundances}\label{subsubsec:ief}

As shown in Table \ref{tab:ia}, the fiducial model starts with H in
molecular form, D in HD, while He, C, N, O are in atomic form.
However, when dense cores form in molecular clouds, a large fraction
of CO and maybe N$_2$ should already be present \citep[see
  also][]{2013A&A...555A..14L}. There is some evidence that a significant fraction
of the nitrogen is still in atomic form in dense cores due to the slow
conversion from N to N$_2$, but the exact amount is unclear
\citep{2010A&A...513A..41H}. As different initial abundances could impact
$D_{\rm frac,eq}^{\rm species}$ and $t_{\rm eq,90}$($D_{\rm frac}^{\rm
  species}$), we quantify these effects considering 3 variations to
the fiducial model descibed in \S\ref{subsec:fiducialmodel}: (1)
``atomic D'', where D is in atomic form, compared to the fiducial
model, assuming that H$_2$ is in molecular form;
(2) ``fully molecular'', where everything starts in molecular form
(all N in N$_2$, all C in CO, with the leftover Oxygen left in atomic
form); (3) ``half N in N$_2$'', where half of the Nitrogen is left in
atomic form compared to ``fully molecular'' case.

\begin{figure}
\epsscale{1.2}
\plotone{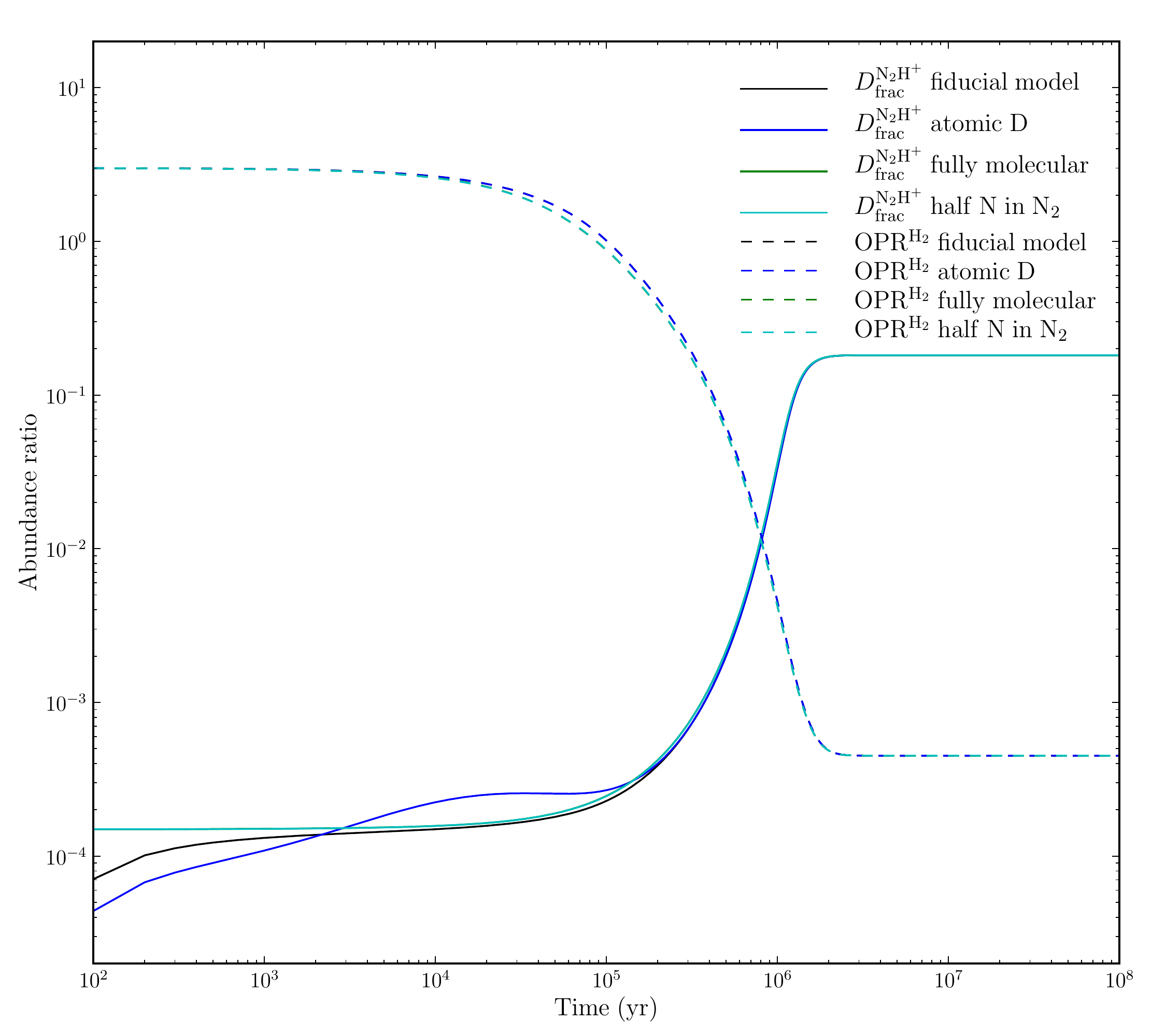}
\caption{
Time evolution of OPR$^{\rm H_2}$ and $D_{\rm frac}^{\rm N_2H^+}$ with
4 sets of initial elemental abundances. See \S\ref{subsubsec:ief} for
the description of these sets. The equilibrium ratios and timescales
are summarized in Table \ref{tab:dfrac}.\label{fig:fam}}
\end{figure}

Figure~\ref{fig:fam} shows the results of these tests, focussing on
the effects on the time evolution of $D_{\rm frac}^{\rm N_2H^+}$ and
OPR$^{\rm H_2}$. Table \ref{tab:dfrac} lists the equilibrium ratios
and timescales. The choice of initial atomic versus molecular
abundances has no effect on the equilibrium abundance ratios and has
little effect ($\lesssim 1\%$) on the timescales.

\begin{deluxetable*}{lcccccc}
\tabletypesize{\scriptsize}
\tablecaption{Equilibrium abundance ratios and timescales for models with various initial conditions.\label{tab:dfrac}}
\tablewidth{0pt}
\tablehead{
\colhead{Model}&\colhead{OPR$_{\rm eq}^{\rm H_2}$}&\colhead{$t_{\rm eq}$(OPR$^{\rm H_2}$)}&\colhead{$t_{\rm eq,90}$(OPR$^{\rm H_2}$)}&\colhead{$D_{\rm frac,eq}^{\rm N_2H^+}$}&\colhead{$t_{\rm eq}$($D_{\rm frac}^{\rm N_2H^+}$)}&\colhead{$t_{\rm eq,90}$($D_{\rm frac}^{\rm N_2H^+}$)}\\
\colhead{}&\colhead{($\times$10$^{-4}$)}&\colhead{(10$^6$ yr)}&\colhead{(10$^6$ yr)}&\colhead{}&\colhead{(10$^6$ yr)}&\colhead{(10$^6$ yr)}
}
\startdata
fiducial & 4.51 & 2.98 & 1.93 & 0.181 & 2.68 & 1.65\\
atomic D & 4.51 & 2.98 & 1.93 & 0.181 & 2.68 & 1.65\\
fully molecular	& 4.51 & 2.98 & 1.92 & 0.181 & 2.66 & 1.63\\
half N in N$_2$	& 4.51 & 2.98 & 1.92 & 0.181 & 2.66 & 1.63\\
\hline
OPR$^{\rm H_2}(t=0)$ = 3 & 4.51 & 2.98 & 1.93 & 0.181 & 2.68 & 1.65\\
OPR$^{\rm H_2}(t=0)$ = 1  & 4.51 & 2.92 & 1.86 & 0.181 & 2.60 & 1.57\\
OPR$^{\rm H_2}(t=0)$ = 0.1 & 4.51 & 2.60 & 1.54 & 0.181 & 2.28 & 1.25\\
OPR$^{\rm H_2}(t=0)$ = 0.01 & 4.51 & 2.18 & 1.12	& 0.181  & 1.86 & 0.830\\
OPR$^{\rm H_2}(t=0)$ = 0.001 & 4.51 & 1.64 & 0.578 & 0.181 & 1.32 & 0.290\\
OPR$^{\rm H_2}(t=0)$ = 0.0007 & 4.51 & 1.48 & 0.429 & 0.181 & 1.16 & 0.152\\
\hline
Maximum $D_{\rm frac}$ model\tablenotemark{a} & 1.33 & 0.725 & 0.429 & 0.903 & 0.645 & 0.338
\enddata
\tablenotetext{a}{See \S\ref{subsec:highestDfrac}.}
\end{deluxetable*}

\subsubsection{Initial OPR$^{\rm H_2}$}\label{subsubsec:iop}

Another poorly constrained, but crucial, parameter is the initial
OPR$^{\rm H_2}$. There are only a few studies yielding observational
constraints: in diffuse clouds, \citet{Crabtree2011} measured
OPR$^{\rm H_2}$ $\simeq$0.3--0.8; in the pre-stellar core L183,
P09 derived OPR$^{\rm H_2}$ $\simeq$0.1 \citep[see
  also][]{2011ApJ...739L..35P}, while \citet{2009A&A...506.1243T} estimated OPR$^{\rm
  H_2}<1$ 
and \citet{2007ApJ...664..956M} estimated OPR$^{\rm H_2}\sim 0.015$ in the starless
Bok globule B68 \citep[see also discussion
  in][S13]{Flower2006}. Evidently, different environmental
conditions strongly impact OPR$^{\rm H_2}$ \citep[as also deduced
  by][in their study of o-H$_2$D$^+$ in star-forming
  regions]{Caselli2008}.

\begin{figure}
\epsscale{1.2}
\plotone{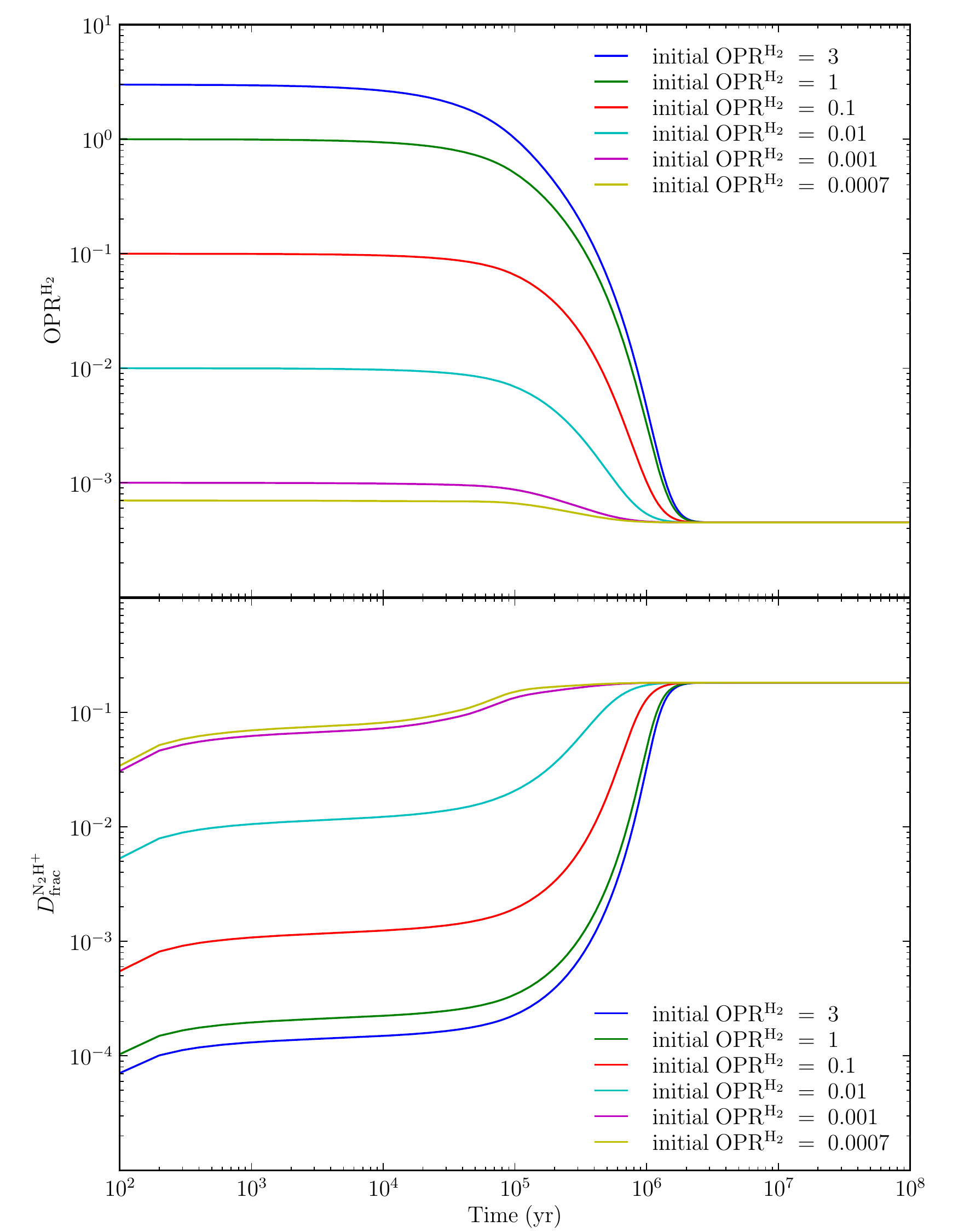}
\caption{
Time evolution of OPR$^{\rm H_2}$ (top) and $D_{\rm frac}^{\rm
  N_2H^+}$ (bottom) under different assumptions of initial OPR$^{\rm
  H_2}$. We explore initial OPR$^{\rm H_2}$ from 3 (the fiducial
model) down to 7$\times$10$^{-4}$. Note that OPR$_{\rm eq}^{\rm H_2}
\simeq 4.51\times 10^{-4}$.\label{fig:on}}
\end{figure}

Our fiducial model starts with OPR$^{\rm H_2}$ = 3, which implies that
all H$_2$ molecules are initially in their statistical spin ratio, as
expected if they have just been formed on the surface of dust grains,
i.e. if the molecular cloud is very young.  However, this may not be
the case, based on the above mentioned observations in diffuse clouds
and if cloud cores form at a later stage compared to the formation of
the parent molecular cloud. To explore this, we consider the effect of
different initial OPR$^{\rm H_2}$ values in the fiducial
model. Fig. \ref{fig:on} shows their effects on the time evolution of
OPR$^{\rm H_2}$ and $D_{\rm frac}^{\rm N_2H^+}$. The different initial
OPR$^{\rm H_2}$ values have little effect on both OPR$_{\rm eq}^{\rm
  H_2}$ and $D_{\rm frac,eq}^{\rm N_2H^+}$, but the timescales to
reach equilibrium are changed \citep[as also found by][see their
  Fig. 2]{2011ApJ...739L..35P}.  Since the OPR$_{\rm eq}^{\rm H_2}$ is
4.51$\times$10$^{-4}$, the lower the initial OPR$^{\rm H_2}$, the
sooner chemical equilibrium will be reached. We find $t_{\rm
  eq,90}$($D_{\rm frac}^{\rm N_2H^+}$) becomes similar to 
$t_{\rm ff}$ if OPR$^{\rm H_2}$ is initially 0.001 or lower. This is
also summarized in Table \ref{tab:dfrac}.  These results suggest that
we should in general consider the possible effects of starting with
much lower values of the initial OPR$^{\rm H_2}$ than the fiducial
value of 3 \citep[although initial values lower than 0.1 are not consistent with 
DCO$^+$ observations, as discussed in][]{2011ApJ...739L..35P}.

\subsection{Effect of Starless Core Physical and Environmental Properties on the Deuteration Timescale}\label{subsec:pe}

Here we present a parameter space exploration to see how different
physical conditions impact the chemical evolution of gas in starless
cores. We will first assume an initial OPR$^{\rm H_2}=3$ (results for
OPR$^{\rm H_2}=1$, 0.1, 0.01 are discussed below in
\S\ref{S:initialOPR}).
Then, we vary four parameters: 
the H number density $n_{\rm H}$ from 10$^3$ to
10$^7$~cm$^{-3}$, the temperature $T$ from 5 to 30~K, the CR
ionization rate $\zeta$ from 10$^{-18}$ to 10$^{-15}$~s$^{-1}$, the
gas phase depletion factor $f_{\rm D}$ from 1 to 1000 ($f_{\rm D}$ = 1 implies
no depletion)\footnote{Note that some small regions of parameter space are not
  internally self-consistent, e.g. a very cold temperature model with
  very high CR ionization rate, but our goal here is to first explore
  the effects of each variable on the deuteration chemistry in
  isolation, before later building self-consistent thermodynamic
  models.}.
These ranges of parameter space are chosen to cover conditions
  expected for both low and high-mass starless cores 
  \citep[e.g.,][]{2007ARA&A..45..339B,2014arXiv1402.0919T}.

\begin{figure*}
\epsscale{1.}
\plotone{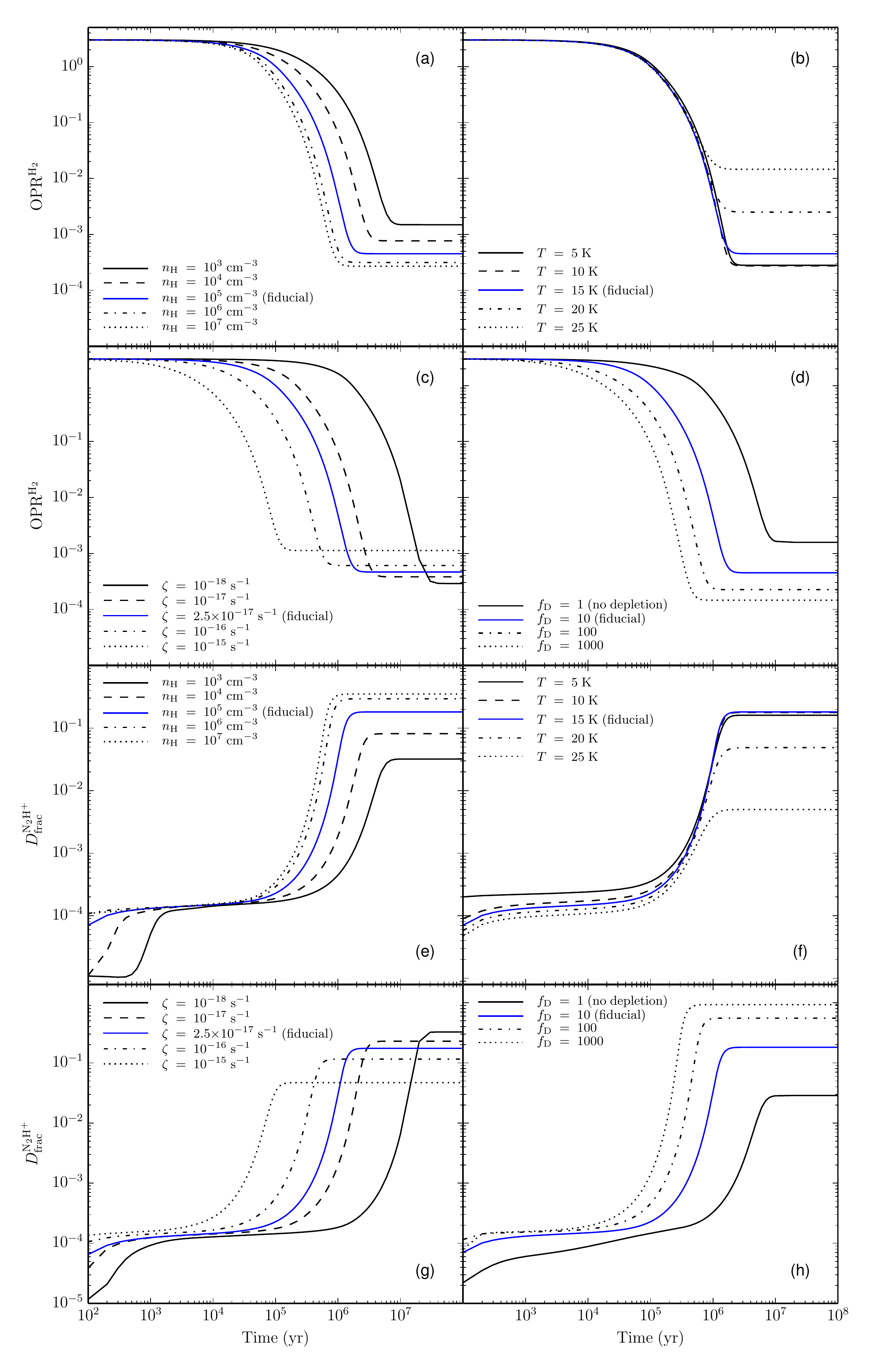}
\caption{
Time evolution of OPR$^{\rm H_2}$ (panels a to d) and $D_{\rm
  frac}^{\rm N_2H^+}$ (panels e to h) for various densities (a, e),
temperatures (b, f), cosmic ray ionization rates (c, g) and depletion
factors (d, h). The blue solid lines correspond to the fiducial model
(as in Fig. \ref{fig:fd}). In each case of exploring the effect of
varying a particular parameter, the other parameter values are set to
those of the fiducial model. See \S\ref{subsec:pe} for the complete
description of the exploration.\label{fig:expld}}
\end{figure*}

Figure~\ref{fig:expld} shows the effect on the time evolution of
OPR$^{\rm H_2}$ and $D_{\rm frac}^{\rm N_2H^+}$ of varying these four
parameters. In general, $n_{\rm H}$ and $T$ have a greater influence
on $D_{\rm frac,eq}^{\rm N_2H^+}$, while $\zeta$ and $f_{\rm D}$
has an effect on both $D_{\rm frac,eq}^{\rm N_2H^+}$ and OPR$^{\rm H_2}$. This
implies that the physical environment plays an important role in dense
core chemistry. It is thus crucial to have good observational
constraints on these properties when trying to model observed cores.

\begin{figure*}
\epsscale{1.18}
\plotone{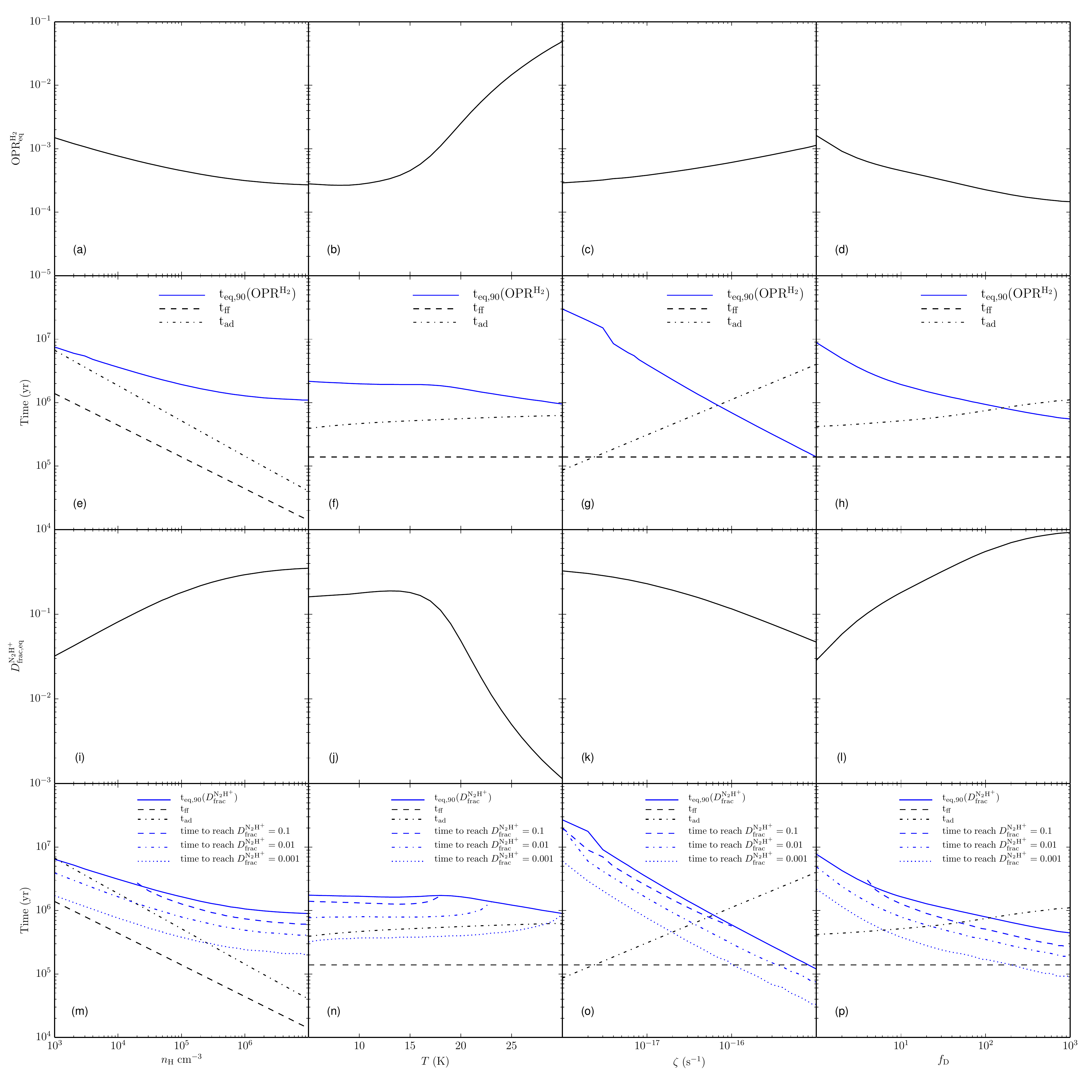}
\caption{
Parameter-space exploration of dependence of 
OPR$_{\rm eq}^{\rm H_2}$ (top row), $t_{\rm eq,90}$(OPR$^{\rm H_2}$)
(2nd row), $D_{\rm frac,eq}^{\rm N_2H^+}$ (3rd row), $t_{\rm
  eq,90}$($D_{\rm frac}^{\rm N_2H^+}$) (bottom row) as a function of
density $n_{\rm H}$ (left column), temperature $T$ (2nd column),
cosmic ray ionization rate $\zeta$ (3rd column), and depletion factor
$f_{\rm D}$ (right column) (see \S\ref{subsec:dfrac} for definitions). In
the 4th row, we also show the times to reach $D_{\rm frac}^{\rm N_2H^+}$ =
0.1, 0.01, 0.001 (missing portions of the lines imply $D_{\rm
  frac}^{\rm N_2H^+}$ does not reach the value of interest for these
conditions).  Also shown are free-fall time $t_{\rm ff}$
(Eq.~\ref{eq:ff}) and ambipolar diffusion time $t_{\rm ad}$
(\S\ref{subsec:tad}) to be compared to $t_{\rm eq,90}$(OPR$^{\rm
  H_2}$) and $t_{\rm eq,90}$($D_{\rm frac}^{\rm N_2H^+}$).\label{fig:exple}}
\end{figure*}

Figure~\ref{fig:exple} shows the variation of the equilibrium ratios
and timescales of OPR$^{\rm H_2}$ (upper 2 rows) and $D_{\rm
  frac}^{\rm N_2H^+}$ (lower 2 rows). Note that the variation of the
equilibrium time of $D_{\rm frac}^{\rm N_2H^+}$ (fourth row) is very
similar to that of OPR$^{\rm H_2}$ (second row), as explained in
\S\ref{subsec:fiducial}.
In the following, we summarize their dependence on each physical
quantity, with emphasis on $D_{\rm frac}^{\rm N_2H^+}$ and $t_{\rm
  eq,90}$($D_{\rm frac}^{\rm N_2H^+}$).

\subsubsection{Dependence on $n_{\rm H}$}\label{subsubsec:nh}

As shown in panel (i) of Figure~\ref{fig:exple}, a denser core will
have a higher $D_{\rm frac,eq}^{\rm N_2H^+}$. 
The $D_{\rm frac,eq}^{\rm N_2H^+}$ changes by 
$\sim$ an order of magnitude, from 3.21$\times$10$^{-2}$ at $n_{\rm H}$
= 10$^{3}$~cm$^{-3}$, to 3.51$\times$10$^{-1}$ at $n_{\rm H}$ =
10$^{7}$~cm$^{-3}$.  

From panel (m) we see that $t_{\rm eq,90}$($D_{\rm frac}^{\rm N_2H^+}$) 
varies with $n_{\rm H}$ by a factor of $\sim$ 7.
Thus, cores with a wide range of densities
have similar deuteration timescales, if other conditions are
fixed. When $n_{\rm H}$ $\ga$ 3$\times$10$^4$~cm$^{-3}$, $t_{\rm
  eq,90}$($D_{\rm frac}^{\rm N_2H^+}$) and $t_{\rm eq,90}$(OPR$^{\rm
  H_2}$) are more than 10 $t_{\rm ff}$ (recall $t_{\rm ff}$ is the
local free-fall time at a given density). Thus highly deuterated
cores, i.e. with $D_{\rm frac}^{\rm N_2H^+}\gtrsim 0.1$, that have
such densities and that also satisfy the other fiducial parameters and
assumed initial conditions would, in the context of the assumption of
constant (or slow) density evolution, need to be ``dynamically old'',
i.e. have existed at the current density for significantly longer than their
local free-fall time. Below, in \S\ref{S:DDE}, we will also
place constraints for such cores in the context of
dynamically-evolving densities.

Panel (m) (and panels n, o, p) also show the ambipolar diffusion
timescale, $t_{\rm ad}$, which is expected to be the relevant collapse
timescale in magnetically subcritical cores. It is always longer than
$t_{\rm ff}$. The ambipolar diffusion timescale is discussed in more
detail in \S\ref{subsec:tad}.

\subsubsection{Dependence on $T$}\label{subsubsec:t}

As shown in panel (j) of Figure \ref{fig:exple}, between 5 and 15~K, the
$D_{\rm frac,eq}^{\rm N_2H^+}$ profile is quite flat. Above 15~K,
$D_{\rm frac, eq}^{\rm N_2H^+}$ drops down by almost 2 orders of
magnitude as $T$ approaches 30 K. The maximum $D_{\rm frac,eq}^{\rm
  N_2H^+}$ of 0.19 is achieved at $T$ = 13~K. The profile of $t_{\rm
  eq,90}$($D_{\rm frac}^{\rm N_2H^+}$) is also quite flat across the
explored temperatures (panel n). We find $t_{\rm eq,90}$($D_{\rm
  frac}^{\rm N_2H^+}$) is always greater than 10 $t_{\rm ff}$ except
for the highest temperatures (as is $t_{\rm eq,90}$(OPR$^{\rm
  H_2}$)). At $T \lesssim $15\,K, OPR$_{\rm eq}^{\rm H_2}$ is well
below 0.001, but it goes up quickly at higher temperatures. Regions
warmer than 20 K, as, for example, gas in the proximity of young
stellar objects,
will then experience an increase of the ortho-to-para H$_2$
ratio and thus a drop in the deuterium fraction
(in agreement with findings by \citealt{Fontani2011} in high-mass
star-forming regions and \citealt{Emprechtinger2009} in low-mass 
star-forming regions).

As stated previously, $D_{\rm frac,eq}^{\rm N_2H^+}$ (as well as the deuterium fraction of other
deuterated species) is controlled by OPR$_{\rm eq}^{\rm H_2}$. 
As $T$ goes down to 13 K, OPR$_{\rm eq}^{\rm H_2}$ drops 
(panel (b) of Figure \ref{fig:exple}), and $D_{\rm frac,eq}^{\rm N_2H^+}$
(panel (j) of Figure \ref{fig:exple}) increases. Below 13 K, both
OPR$_{\rm eq}^{\rm H_2}$ and $D_{\rm frac,eq}^{\rm N_2H^+}$
become roughly constant. 
This is because the H$_3^+$ + H$_2$ reacting system contains 
reactions with activation energies that convert p-H$_2$ to o-H$_2$, 
such as:
\begin{equation}\label{equ:destrph26}
\textrm{o-H}_3^+~+~\textrm{p-H}_2~\rightarrow~\textrm{p-H}_3^+~+~\textrm{o-H}_2
\end{equation}
\begin{equation}\label{equ:destrph27}
\textrm{o-H}_3^+~+~\textrm{p-H}_2~\rightarrow~\textrm{o-H}_3^+~+~\textrm{o-H}_2.
\end{equation}
The activation energy barrier of the above reactions are of the order of 100 K. 
As the temperature drops, the endothermic channels are effectively 
closed off and the abundances of both p-H$_2$ and o-H$_2$ 
are controlled mainly by $T$-independent reactions, e.g.,
\begin{equation}\label{equ:destrph23529}
\textrm{p-H}_2^+~+~\textrm{p-H}_2~\rightarrow~\textrm{p-H}_3^+~+~\textrm{H}.
\end{equation} 
Consequently, the OPR$_{\rm eq}^{\rm H_2}$ and $D_{\rm frac,eq}^{\rm N_2H^+}$ 
become almost independent of $T$.

\begin{figure}
\epsscale{1.2}
\plotone{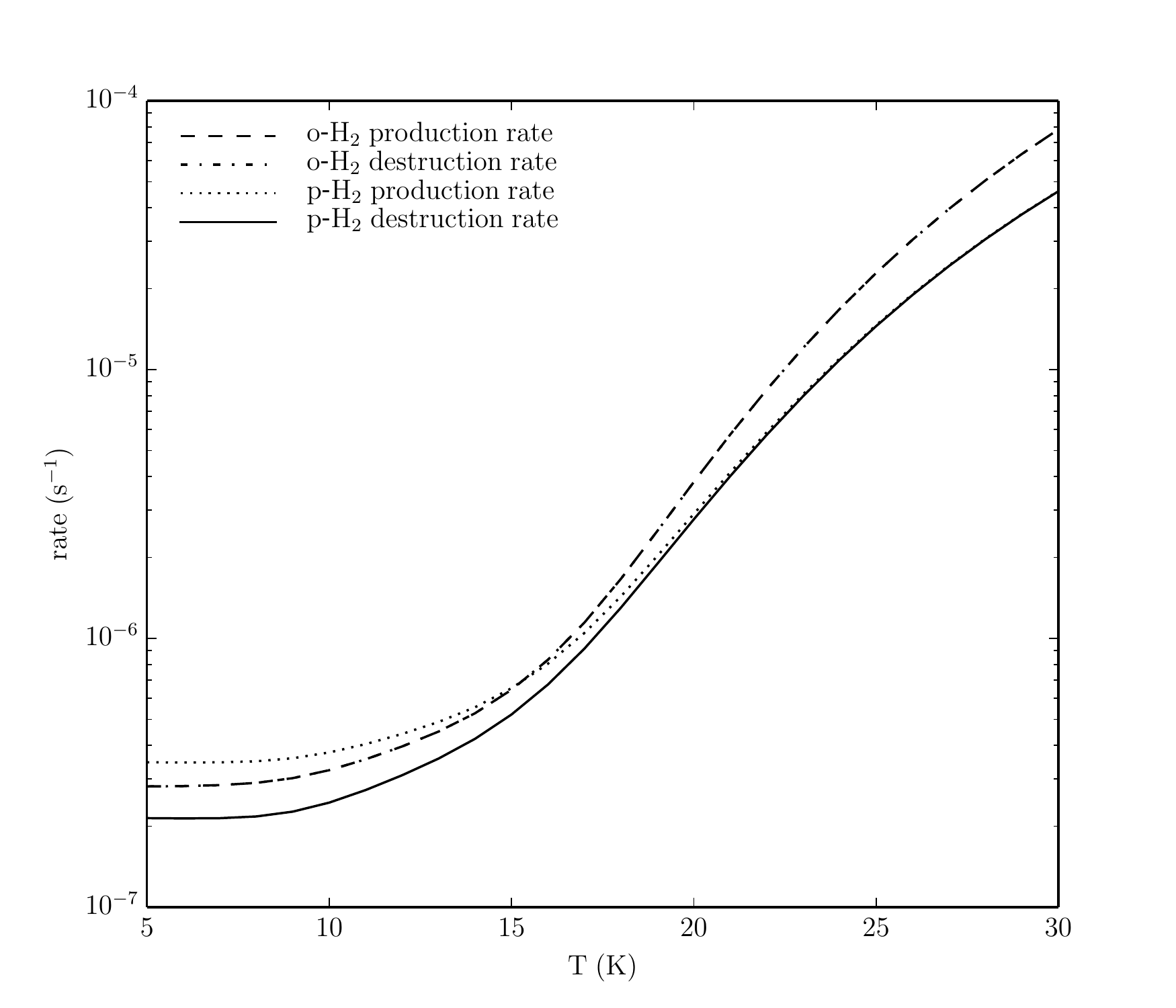}
\caption{
Production and destruction rates of o-H$_2$
and p-H$_2$ at equilibrium time step as a function of T. 
See \S\ref{subsubsec:t}.\label{fig:ph2}}
\end{figure}

\subsubsection{Dependence on $\zeta$}

$D_{\rm frac,eq}^{\rm N_2H^+}$ drops by a factor of 6
as $\zeta$ increases from $10^{-18}$ to $10^{-15}\:{\rm
  s}^{-1}$ (panel k), due to the enhanced electron abundance and the
consequent dissociative recombination of the deuterated isotopologues
of H$_3^+$ \citep[see also][]{Caselli2008}. The highest $D_{\rm
  frac,eq}^{\rm N_2H^+}$ of 0.33 appears at the lowest $\zeta$ =
10$^{-18}$~s$^{-1}$.  This shows the importance for the astrochemical
modeling of constraining $\zeta$.  Panel (o) shows that $t_{\rm
  eq,90}$($D_{\rm frac}^{\rm N_2H^+}$) changes by more than 2 orders of
magnitude within the $\zeta$ range explored. The smallest $t_{\rm
  eq,90}$($D_{\rm frac}^{\rm N_2H^+}$) is only 1.2$\times$10$^5$~yr
at $\zeta$ = 10$^{-15}$~s$^{-1}$, which is much shorter than $t_{\rm
  ff}$. 
However, such high CR ionization rates are not expected to be relevant
in typical Galactic star-forming regions (and would also be expected
to yield relatively high kinetic temperatures).
With moderate $\zeta$ ($\lesssim$ 10$^{-16}$~s$^{-1}$), $t_{\rm
  eq,90}$($D_{\rm frac}^{\rm N_2H^+}$) is significantly greater than
$t_{\rm ff}$ ($\gtrsim$ 7 $t_{\rm ff}$) at the fiducial density.
The dependence of OPR$^{\rm H_2}$ with $\zeta$ is shown in panel (c) and
its equilibrium timescale in panel (g).

\subsubsection{Dependence on $f_{\rm D}$}\label{subsubsec:fd}

Panel (l) of Figure~\ref{fig:exple} shows that $D_{\rm frac,eq}^{\rm
  N_2H^+}$ goes up by more than an order of magnitude as $f_{\rm D}$
increases from 1 to 1000. This agrees with the expectation that
depletion of neutral species, in particular CO and O, the main
destruction partners of H$_3^+$ and its deuterated forms, will result
in the enhancement of $D_{\rm frac}^{\rm N_2H^+}$ \citep[see
  also][]{DL1984}. At $f_{\rm D}$ = 1000 we encounter the highest $D_{\rm
  frac,eq}^{\rm N_2H^+}$ = 0.93 in our exploration. Such high values
have seldom been observed \citep[e.g.,][]{2012A&A...538A.137M}. 
We find $t_{\rm eq,90}$($D_{\rm frac}^{\rm N_2H^+}$) decreases with
stronger depletion, which is shown in panel (p), although $t_{\rm
  eq,90}$($D_{\rm frac}^{\rm N_2H^+}$) is at least a factor 7 larger
than $t_{\rm ff}$ when $f_{\rm D}$ $\lesssim$100 at the fiducial density.

\begin{figure}
\epsscale{1.}
\plotone{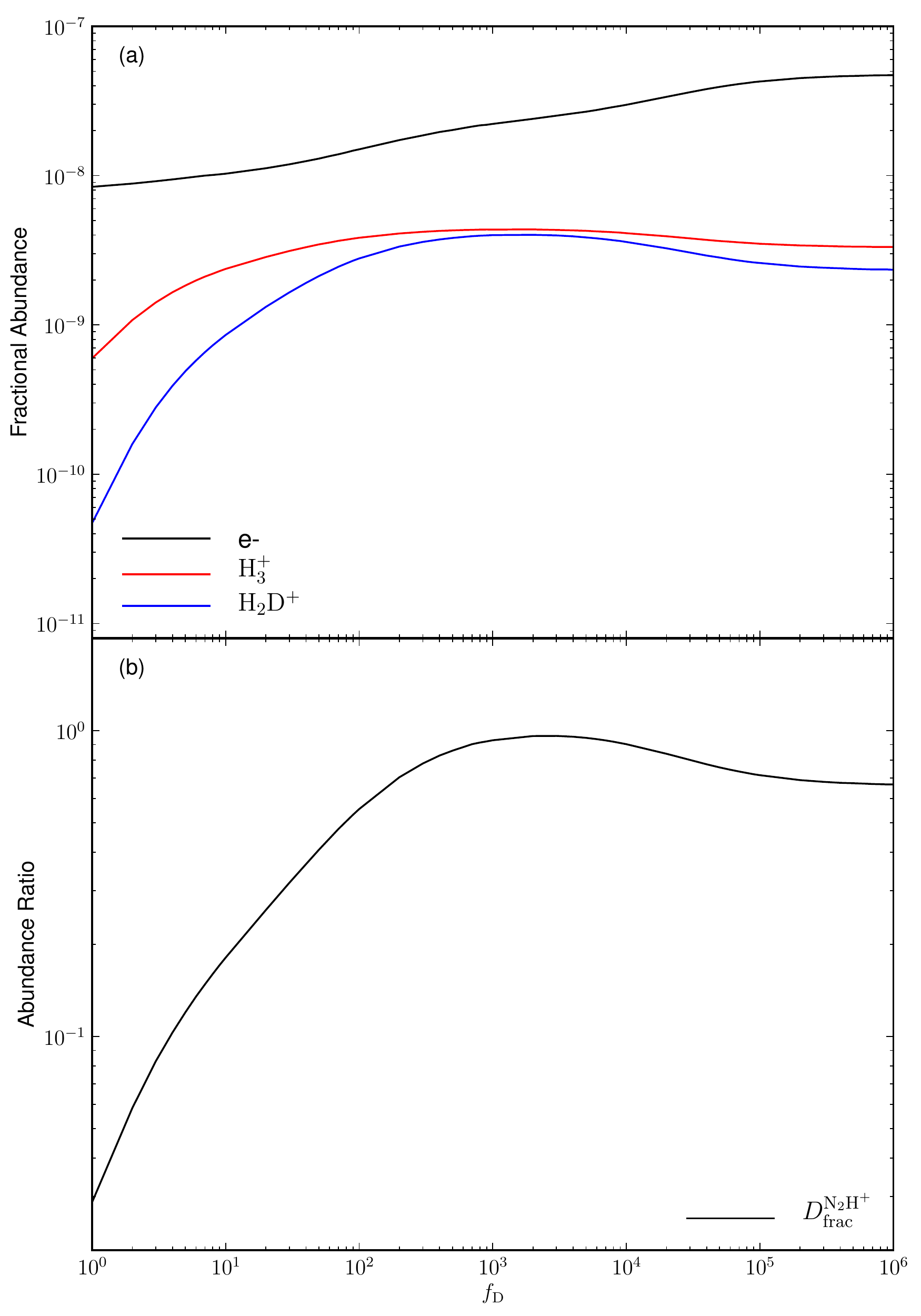}
\caption{
(a) Relations between the depletion factor and the fractional abundances (relative to $n_{\rm H}$) of ${\rm H_3^+}$, ${\rm H_2D^+}$, and electron, respectively. The abundances are taken at the equilibrium step of $D_{\rm frac,eq}^{\rm N_2H^+}$. (b) Same as the panel (l) in Fig. \ref{fig:exple}, but extended to $f_{\rm D}$ = 10$^6$.\label{fig:checkfdbig}}
\end{figure}

\begin{figure}
\epsscale{1.}
\plotone{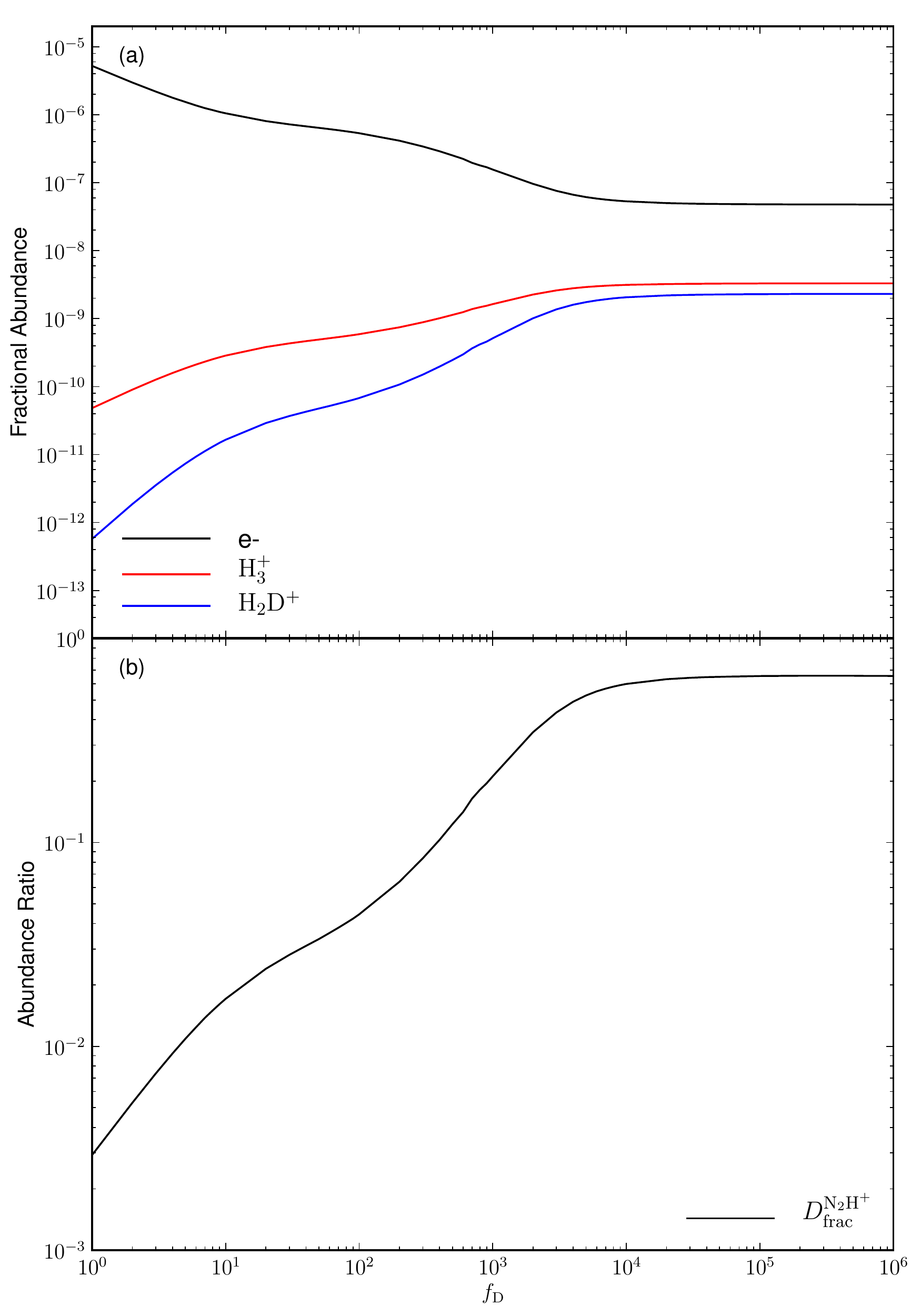}
\caption{
Same as Fig. \ref{fig:checkfdbig}, but now all dissociative recombination reactions between electron and 
O-bearing species (containing no Nitrogen or Carbon) are removed ($\sim$ 40 reactions).
This is used to prove that the "bump" in panel (b) of Fig. \ref{fig:checkfdbig} is caused by 
these reactions (no "bump" after removing the reactions). See \S\ref{subsubsec:fd}.
\label{fig:checkfdelecnoc}}
\end{figure}

An interesting point is whether $D_{\rm frac,eq}^{\rm N_2H^+}$ will 
keep going up with stronger depletion.
We extend our exploration to $f_{\rm D}$ = 10$^6$, 
which is shown in Fig. \ref{fig:checkfdbig}. 
We can see in panel (b), the $D_{\rm frac,eq}^{\rm N_2H^+}$ - $f_{\rm D}$ 
relation drops moderately at $f_{\rm D}$ $\sim$ 2000. 
In panel (a), [${\rm H_3^+}$] and [${\rm H_2D^+}$] reach the peak at 
$f_{\rm D}$ $\sim$ 2000 and drop moderately until $f_{\rm D}$ = 10$^6$. 
Besides the destruction partners like CO, 
electrons can also destroy H$_3^+$ and its deuterated forms. 
We plot the electron abundance versus $f_{\rm D}$ in Fig. \ref{fig:checkfdbig}(a).
As we can see, the electron abundance increases 
at $f_{\rm D}$ $\ga$ 2000 while $D_{\rm frac,eq}^{\rm N_2H^+}$ drops. 
This supports our expectation that the super-depletion of heavy 
elements reduces the destruction partners of electron, 
so that [e-] can approach a high level which suppresses the 
abundances of ${\rm H_3^+}$ and ${\rm H_2D^+}$, etc. 
To confirm this, we remove all dissociative recombination reactions 
between electron and O-bearing species ($\sim$ 40 reactions. 
These species contain no Nitrogen or Carbon)\footnote{The 
reason we choose O-bearing species is that we have made 
another three explorations where we reduce initial [C], [N], [O] 
independently. We denote the depletion of C, N, O with 
$f_{\rm D}$(C), $f_{\rm D}$(N), $f_{\rm D}$(O), respectively. 
We find from the explorations that reducing initial [O] can 
reproduce the drop of $D_{\rm frac,eq}^{\rm N_2H^+}$. 
So we expect that O-bearing species play the crucial role.} and perform 
the exploration again. The results are shown in 
Fig. \ref{fig:checkfdelecnoc}. Now we see that the bump of 
$D_{\rm frac,eq}^{\rm N_2H^+}$ around $f_{\rm D}$ $\sim$ 2000 is gone. 
$D_{\rm frac,eq}^{\rm N_2H^+}$ simply increases with $f_{\rm D}$ 
and reaches a constant value ($\sim$ 0.7). In panel (a) of 
Fig. \ref{fig:checkfdelecnoc}, because of the reduced number 
of dissociative reactions, the electron abundance is high at moderate 
depletion, as compared to Fig. \ref{fig:checkfdbig}(a). 
The [e-] and $D_{\rm frac,eq}^{\rm N_2H^+}$ at extreme $f_{\rm D}$ 
approach the same values as those in Fig. \ref{fig:checkfdbig}, 
respectively. These imply that what we see in panel (b) of 
Fig. \ref{fig:checkfdbig} is the result of the competition between two 
mechanisms: (1) species like CO can destroy H$_3^+$ and its 
deuterated forms; (2) O-bearing species consume electron through 
dissociative reactions. It turns out that in our exploration of $f_{\rm D}$ 
(panel b of Fig. \ref{fig:checkfdbig}), mechanism (1) dominates 
at $f_{\rm D}$ $\la$ 2000, and mechanism (2) dominates at 2000 
$\la$ $f_{\rm D}$ $\la$ 10000. At $f_{\rm D}$ $\ga$ 10000, there are
too few O-bearing species consuming electron, explaining the 
$D_{\rm frac,eq}^{\rm N_2H^+}$ - $f_{\rm D}$ relation in
Fig. \ref{fig:checkfdelecnoc}(b). 

\subsubsection{Dependence on Initial OPR$^{\rm H_2}$}\label{S:initialOPR}

\begin{figure*}
\epsscale{1.18}
\plotone{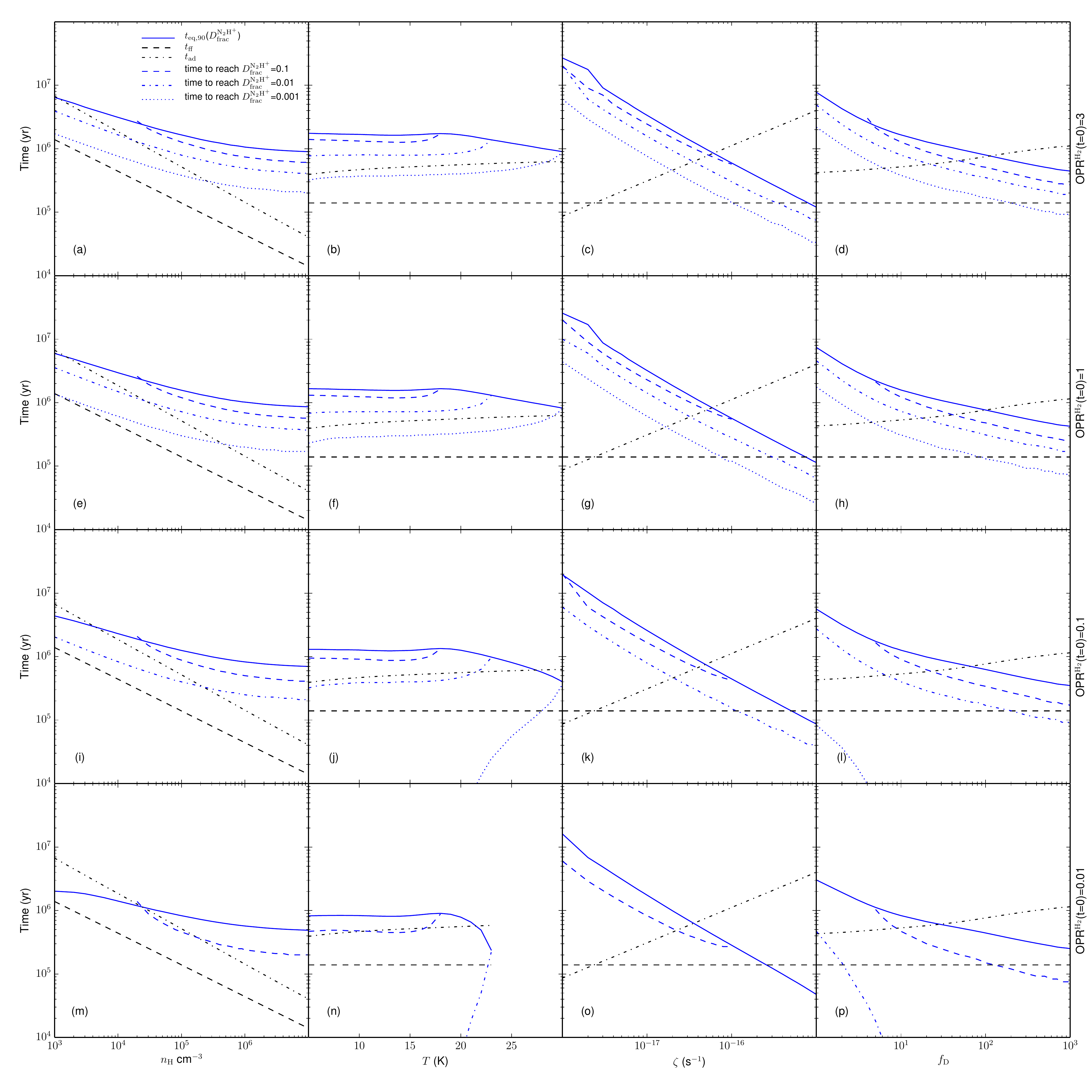}
\caption{
Same as bottom row of Fig. \ref{fig:exple}, which now appears as the
top row here. Then the 2nd, 3rd and bottom rows show the effect of
changing the initial OPR$^{\rm H_2}$ to 1, 0.1 and 0.01,
respectively. The blank parts in high-temperature exploration of the
bottom row are due to this initial OPR$^{\rm H_2}$=0.01 being smaller
than OPR$_{\rm eq}^{\rm H_2}$ (see panel (b) in Fig. \ref{fig:exple}).\label{fig:exple_op}}
\end{figure*}

The effect of varying the initial OPR$^{\rm H_2}$ on the time evolution of
the fiducial model was discussed above in \S\ref{subsubsec:iop}. In
Figure \ref{fig:exple_op}
we show the effect on the deuteration timescale parameter space ($n_{\rm H}$, $T$,
$\zeta$, $f_{\rm D}$) exploration of starting with
OPR$^{\rm H_2}$ = 1, 0.1, 0.01, rather than 3. 

In general, the effect of a lower starting OPR$^{\rm H_2}$ value is to
reduce the timescales needed to reach a given level of $D_{\rm
  frac}^{\rm N_2H^+}$. However, for most of the parameter space, $t_{\rm
  eq,90}$($D_{\rm frac}^{\rm N_2H^+}$) still remains significantly greater
than $t_{\rm ff}$.

\subsubsection[]{Highest $D_{\rm frac}^{\rm N_2H^+}$ predicted in our model}\label{subsec:highestDfrac}

High values of $D_{\rm frac}^{\rm N_2H^+}$ have been reported in
recent observations of starless cores. \citet{Fontani2011} observed
several potential massive starless cores, finding a highest $D_{\rm
  frac}^{\rm N_2H^+}$ = 0.7 in their source Infrared Dark Cloud G2.
An even higher value of $D_{\rm frac}^{\rm N_2H^+}$ = 0.99 has been
reported by \citet[][though this may be affected by the uncertainties
from treating ${\rm N_2H^+}$ with non-LTE model but ${\rm N_2D^+}$
with LTE model]{2012A&A...538A.137M}, toward Orion B9 SMM1. Such high
values are not predicted by our fiducial model. However, it is
interesting if we combine the explored parameters $n_{\rm H}$, $T$,
$\zeta$, $f_{\rm D}$ at the values where $D_{\rm frac,eq}^{\rm N_2H^+}$
peaks ($n_{\rm H}$ = 10$^7$~cm$^{-3}$, $T$ = 13~K, $\zeta$ =
10$^{-18}$~s$^{-1}$, $f_{\rm D}$ = 1000) to gauge the maximum level
of deuteration that can result from our model (might not be global in core).  
Equilibrium ratios and timescales are summarized in Table
\ref{tab:dfrac}. We find $D_{\rm frac}^{\rm N_2H^+}$ goes up to 0.903,
while $t_{\rm eq,90}$($D_{\rm frac}^{\rm N_2H^+}$) is about 46 $t_{\rm
  ff}$ (1.39$\times$10$^4$ yr at $n_{\rm H}$ = 10$^7$~cm$^{-3}$). 
Detailed constraints on the parameter space needed for
individual observed sources will be presented in a future study.

\subsection{Effect of Time-Dependent Depletion/Desorption}\label{S:TDD}

\begin{figure}
\epsscale{1.15}
\plotone{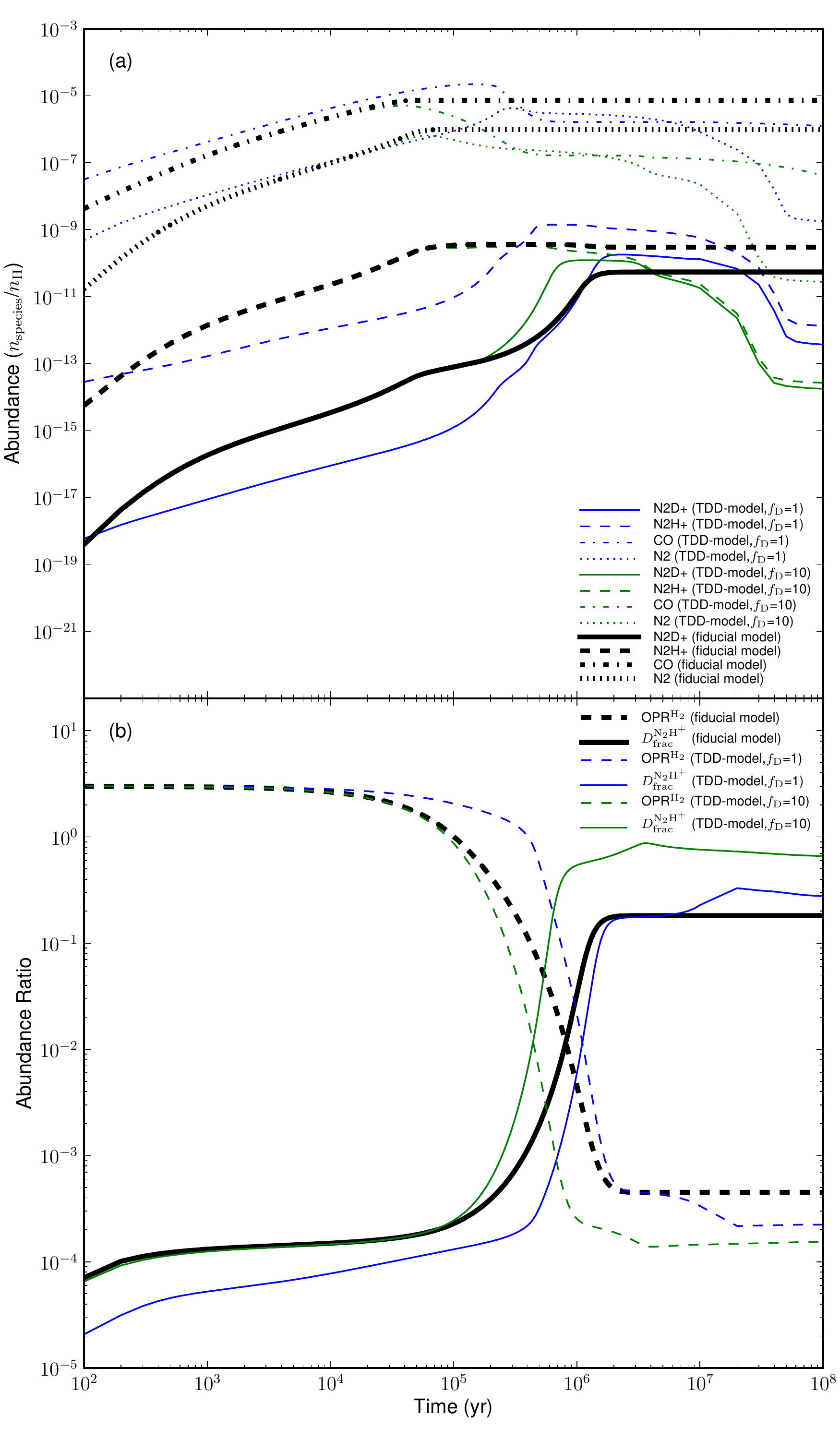}
\caption{
(a) Top panel: Time evolution of fractional abundances of important
   gas-phase species in both the fiducial model (thick black lines)
   and the TDD models (thin lines). The two TDD models shown here started with $f_{\rm D}$=1 (blue) and $f_{\rm D}$=10 (green).
(b) Bottom panel: Time evolution of gas-phase OPR$^{\rm H_2}$ and
   $D_{\rm frac}^{\rm N_2H^+}$ in the fiducial model (thick black
   lines) and the TDD model (thin lines). The two TDD models shown here started with $f_{\rm D}$=1 (blue) and $f_{\rm D}$=10 (green).\label{fig:fddep}}
\end{figure}

In Figure \ref{fig:fddep} we compare the fiducial model (with constant
$f_{\rm D}$=10) and the TDD model (with starting values of $f_{\rm D}$=1 and
$f_{\rm D}$=10). Panel (a) shows the time evolution of the fractional
abundances of ${\rm N_2D^+}$, ${\rm N_2H^+}$, CO, and ${\rm N_2}$.
The time evolution of the abundance of these species show
qualitatively similar behaviours in the two models, with 
modest quantitative differences. We note that the TDD models do not
reach equilibrium within $10^8$~yr because of continuing freeze-out,
especially of $\rm N_2$. The ${\rm N_2D^+}$ abundance shows a plateau
between 5$\times$10$^6$ and 5$\times$10$^7$ yr, before dropping
together with the ${\rm N_2}$ abundance.

In panel (b) we compare gas-phase OPR$^{\rm H_2}$ and $D_{\rm
  frac}^{\rm N_2H^+}$ between the fiducial model and the TDD models.
Compared to the fiducial model, the decline of OPR$^{\rm H_2}$ is
slower in the $f_{\rm D}$=1 TDD model and faster in the $f_{\rm D}$=10 TDD model,
so that for most of the time evolution, up to $\sim 10^7$~yr, the
fiducial models results are bracketed by the TDD models. 

Note our simple TDD models do not include surface chemistry, since
this opens up even larger uncertainties, which we defer to a future
study. As an initial check to see if surface chemistry can have a
significant effect, we have examined the
S13 models with and without surface reactions. The effect 
of including surface chemistry within these models on OPR$^{\rm H_2}$
is very minor, and the time evolution of $D_{\rm frac}^{\rm
  N_2H^+}$ is also largely unaffected.
Thus we do not expect our fiducial or TDD model results to be
significantly affected by neglect of surface chemistry. 

\subsection[]{Effect of Dynamical Density Evolution} \label{S:DDE}

We have so far presented models that treat density as an unchanging,
controllable parameter. Here we carry out a set of Dynamical Density
Evolution (DDE) models that examine various rates of collapse relative
to the free-fall rate by which a core of current density $n_{\rm H,1}$
at current time $t_1$ is created from a core at starting density
$n_{\rm H,0}$ at starting time $t_0$. We parameterize the rate of
density increase via
\begin{equation}\label{equ:denschangerate}
\frac{{\rm d}n_{\rm H}}{{\rm d}t} = \alpha_{\rm ff}\frac{n_{\rm H}(t)}{t_{\rm ff}(t)}
\end{equation}
where $t_{\rm ff}$ is the local free-fall time at current density
$n_{\rm H}$ (Eq. \ref{eq:ff}) and $\alpha_{\rm ff}$ is a parameter
controlling how fast the core collapses. We define a past time
variable that increases going back into a core's history via
\begin{equation}\label{eq:tpast}
t_{\rm past} = t_1 - t.
\end{equation}
So the past density evolution is described by
\begin{equation}\label{equ:timedens}
n_{\rm H,past} = n_{\rm H,1}\left[ 1 + 3.60 \alpha_{\rm ff} \left( \frac{n_{\rm H,1}}{10^5\:{\rm cm^{-3}}}\right)^{1/2} \left(\frac{t_{\rm past}}{10^6\:{\rm yr}}\right)\right]^{-2}.
\end{equation}

\begin{figure*}
\epsscale{1.15}
\plotone{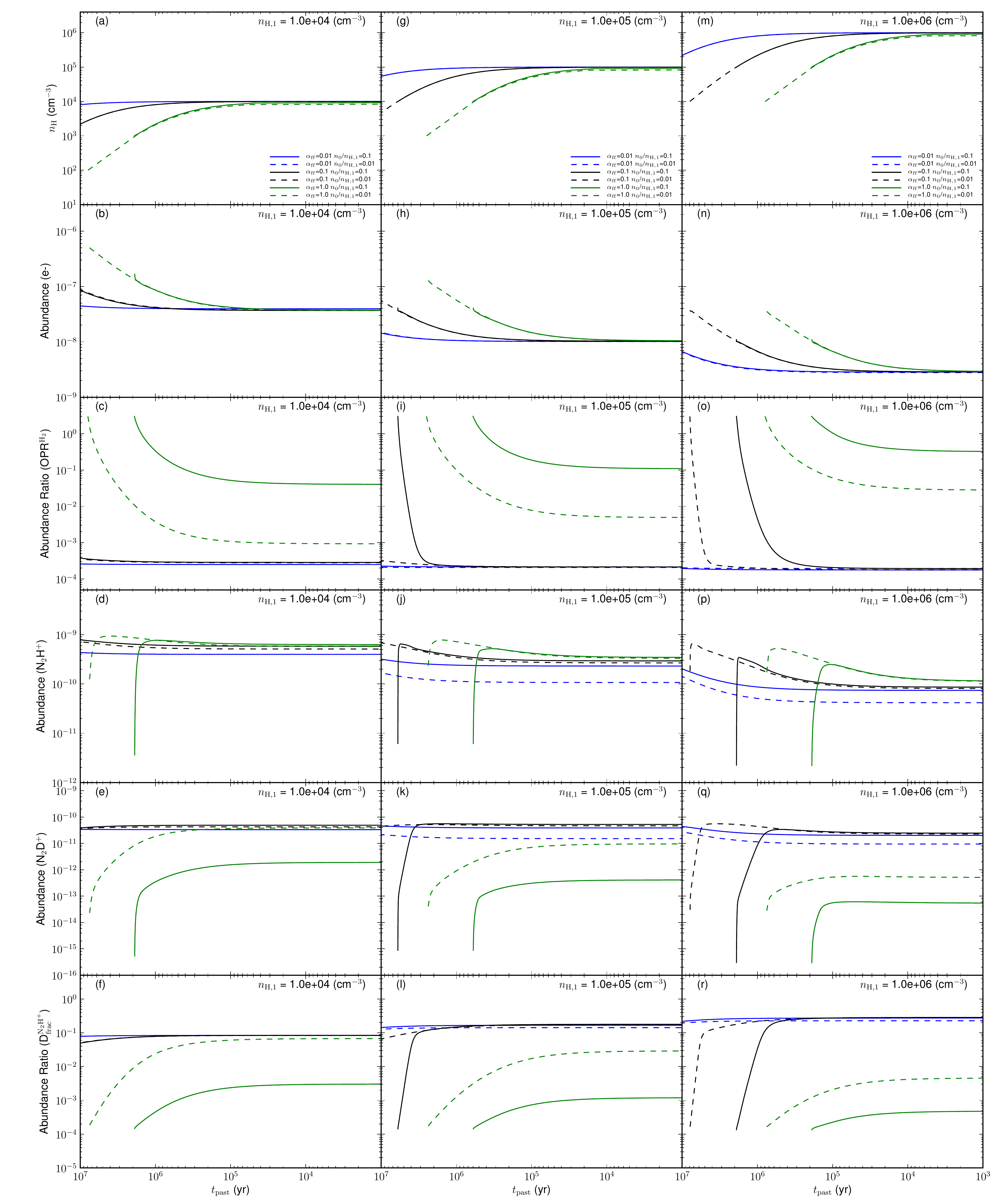}
\caption{
Dynamical Density Evolution (DDE) models that have a time-evolving density
at various rates relative to free-fall collapse, as parameterized by
$\alpha_{\rm ff}$ (see Eq. \ref{equ:timedens}). Each column shows the
results of particular target densities $n_{\rm H,1} = 10^4,10^5,
10^6\:{\rm cm^{-3}}$ (left to right). The top row shows the time
evolution of the density as a function of $t_{\rm past}$, increasing
to the left. In each case, models with $\alpha_{\rm ff}=0.01,0.1,1$
and starting to final density ratios of $n_{\rm H,0}/n_{\rm H,1}=0.1,
0.01$ are shown.
Then, rows 2-6 show the time evolution of [e-], OPR$^{\rm H_2}$,
[${\rm N_2H^+}$], [${\rm N_2D^+}$], $D_{\rm frac}^{\rm N_2H^+}$,
respectively.\label{fig:dynam_gas}}
\end{figure*}

For a given current ``target'' density, $n_{\rm H,1}$, we then explore
different ratios of starting density: $n_{\rm H,0}/n_{\rm
  H,1}=0.1, 0.01$ and three different values of $\alpha_{\rm
  ff} = 0.01, 0.1, 1$. We run these models for three different target
densities $n_{\rm H,1} = 10^4, 10^5, 10^6\:{\rm cm^{-3}}$. 
We first start by keeping other aspects of the modeling the same as
the fiducial model, i.e. a starting OPR$_{\rm eq}^{\rm H_2}$ = 3 and a
fixed depletion factor of $f_{\rm D}$=10. The results are shown in Figure
\ref{fig:dynam_gas}.

The first row of Figure \ref{fig:dynam_gas} shows the density
evolution with $t_{\rm past}$ increasing to the left. For each $n_{\rm
  H,1}$, the faster the collapse rate (larger $\alpha_{\rm ff}$), the
shorter the past history of the core since its starting condition.
Similarly, for fixed $n_{\rm H,1}$ and $\alpha_{\rm ff}$, larger
values of $n_{\rm H,0}$ mean shorter core histories. 
The second row of Figure \ref{fig:dynam_gas} shows the evolution of
the ionization fraction, which declines as density increases. The
third row shows the evolution of OPR$_{\rm eq}^{\rm H_2}$, showing
rapid falls from the assumed starting value of 3. Note that in some of
the fast-evolving, higher density models there is insufficient time
for OPR$_{\rm eq}^{\rm H_2}$ to reach its equilibrium value.
The fourth and fifth rows show the abundances of $\rm N_2H^+$ and 
$\rm N_2D^+$, respectively, while the bottom row shows the evolution of
$D_{\rm frac}^{\rm N_2H^+}$. Again, note that in the fast-evolving,
higher density models there is insufficient time to reach $D_{\rm frac,eq}^{\rm N_2H^+}$.

Note that very slowly evolving models with $t_{\rm past}$ extending
beyond several $\times 10^7$~yr are unlikely to be relevant given
estimated GMC lifetimes \citep[e.g., $\sim 3\times 10^7$~yr,][]{1997ApJ...476..166W}.
Considering the cases with the fastest collapse with $\alpha_{\rm
  ff}=1$ that create cores with $n_{\rm H} = 10^5$ to $10^6\:{\rm
  cm^{-3}}$ from starting conditions a factor of 10 lower in density
(green solid lines in panels (i), (o), (l), (r) of
Fig. \ref{fig:dynam_gas}), then the collapse history did not produce
very low OPR$^{\rm H_2}$ or very high $D_{\rm frac}^{\rm N_2H^+}$
(always $< 10^{-3}$). However, models of fast collapse could
potentially form highly deuterated cores if starting from lower
densities (thus giving more time for chemical evolution) or, as
explored below, with lower initial OPR$_{\rm eq}^{\rm H_2}$ ratios.

\begin{figure*}
\epsscale{1.15}
\plotone{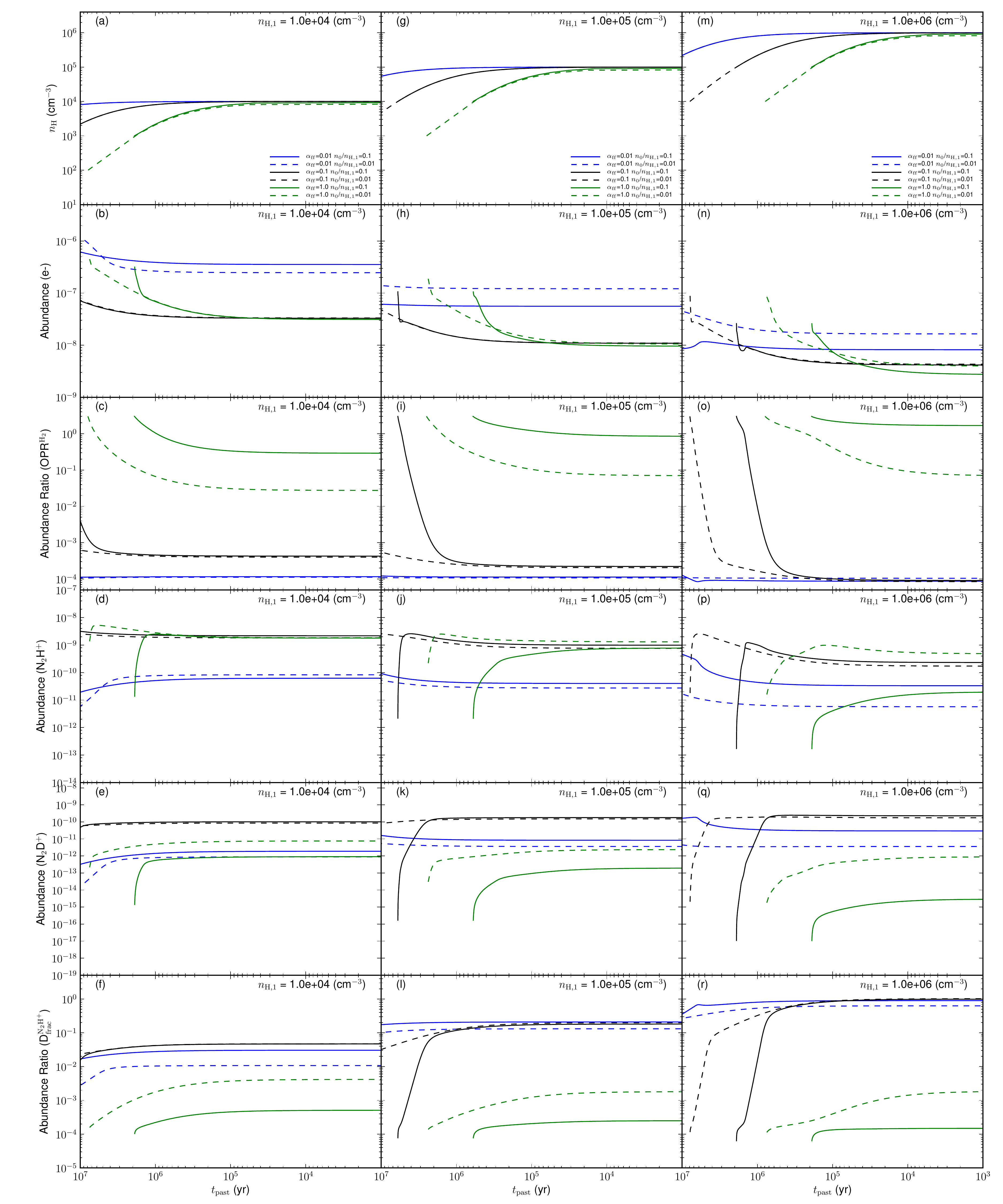}
\caption{
Same as Fig. \ref{fig:dynam_gas}, but all models are now with
Time-Dependent Depletion/Desorption (TDD).\label{fig:dynam_grain}}
\end{figure*}

We next re-run the above DDE models, but with time-dependent
depletion/desorption starting from $f_{\rm D}$=1. These TDD+DDE models are
shown in Figure~\ref{fig:dynam_grain}. We find broadly similar results
that rapidly collapsing high density cores have difficulty achieving
high levels of $D_{\rm frac}^{\rm N_2H^+}$.

\begin{figure*}
\epsscale{0.9}
\plotone{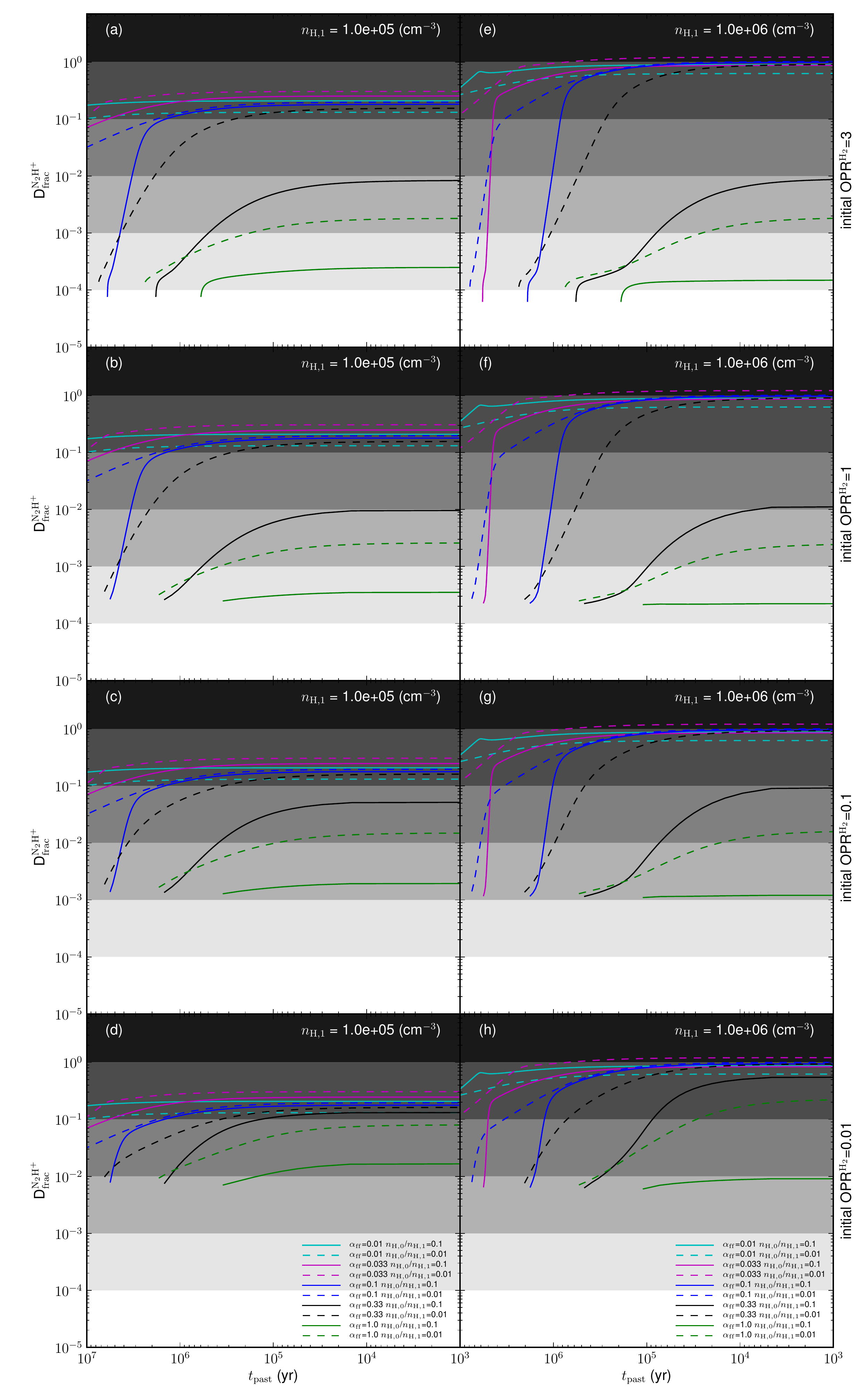}
\caption{
Effect of starting OPR$^{\rm H_2}$ on $D_{\rm frac}^{\rm N_2H^+}$ in
dynamical density evolution with time-dependent depletion/desorption
(DDE+TDD) models of dense cores. Left and right columns show the
results of target densities $n_{\rm H,1} = 10^5, 10^6\:{\rm cm^{-3}}$,
respectively. From top to bottom, the rows show starting OPR$^{\rm
  H_2}$=3, 1, 0.1, 0.01, respectively. In each case, models with
$\alpha_{\rm ff}=0.01, 0.033, 0.1, 0.33, 1$ and starting to final
density ratios of $n_{\rm H,0}/n_{\rm H,1}=0.1, 0.01$ are shown.
Here the starting $f_{\rm D}$=1.\label{fig:dynam_grain_op}}
\end{figure*}

\begin{figure*}
\epsscale{.9}
\plotone{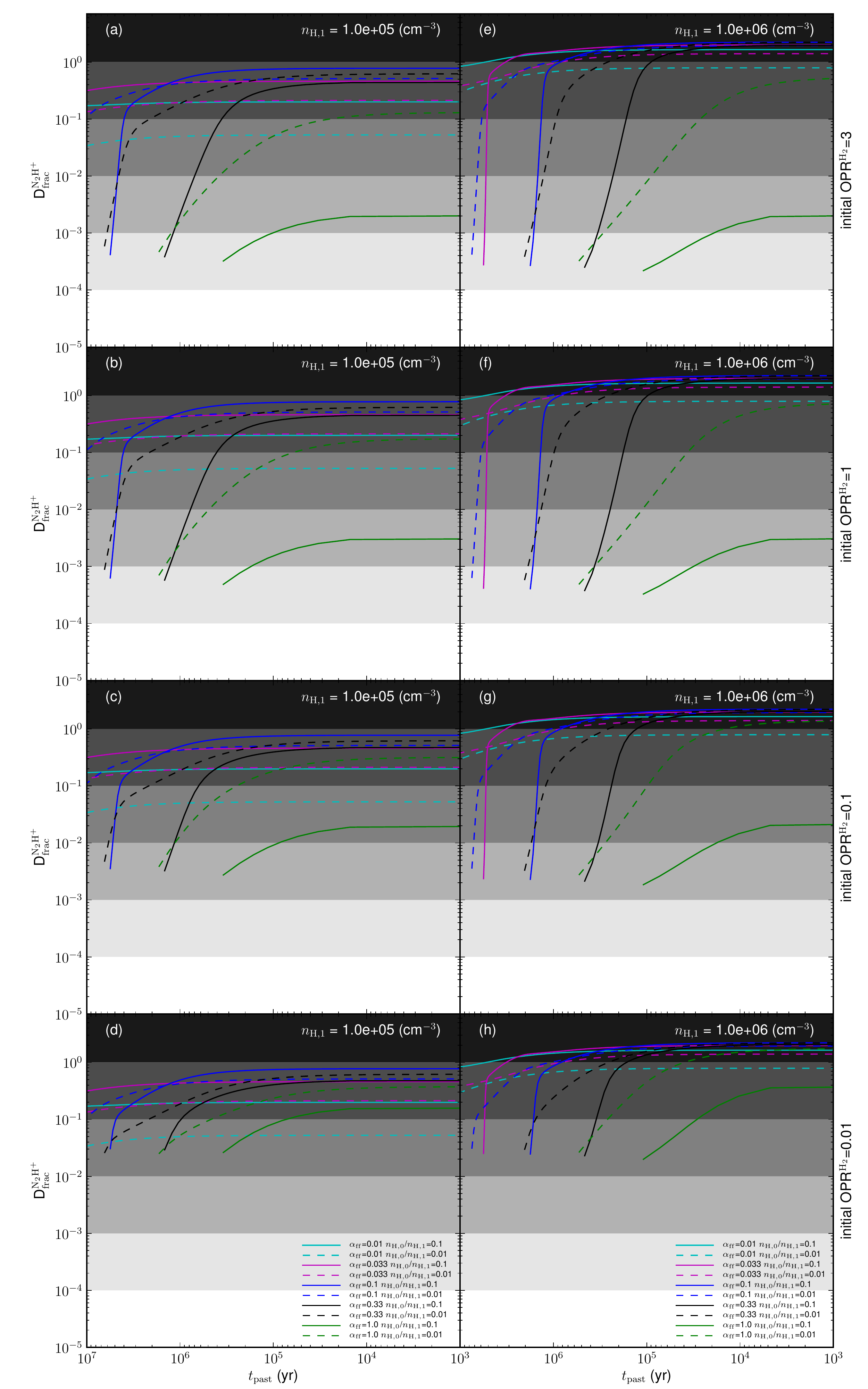}
\caption{
Same as Fig. \ref{fig:dynam_grain_op} but starting with $f_{\rm D}$=10.\label{fig:dynam_grain_op_fd10}}
\end{figure*}

\begin{figure*}
\epsscale{.9}
\plotone{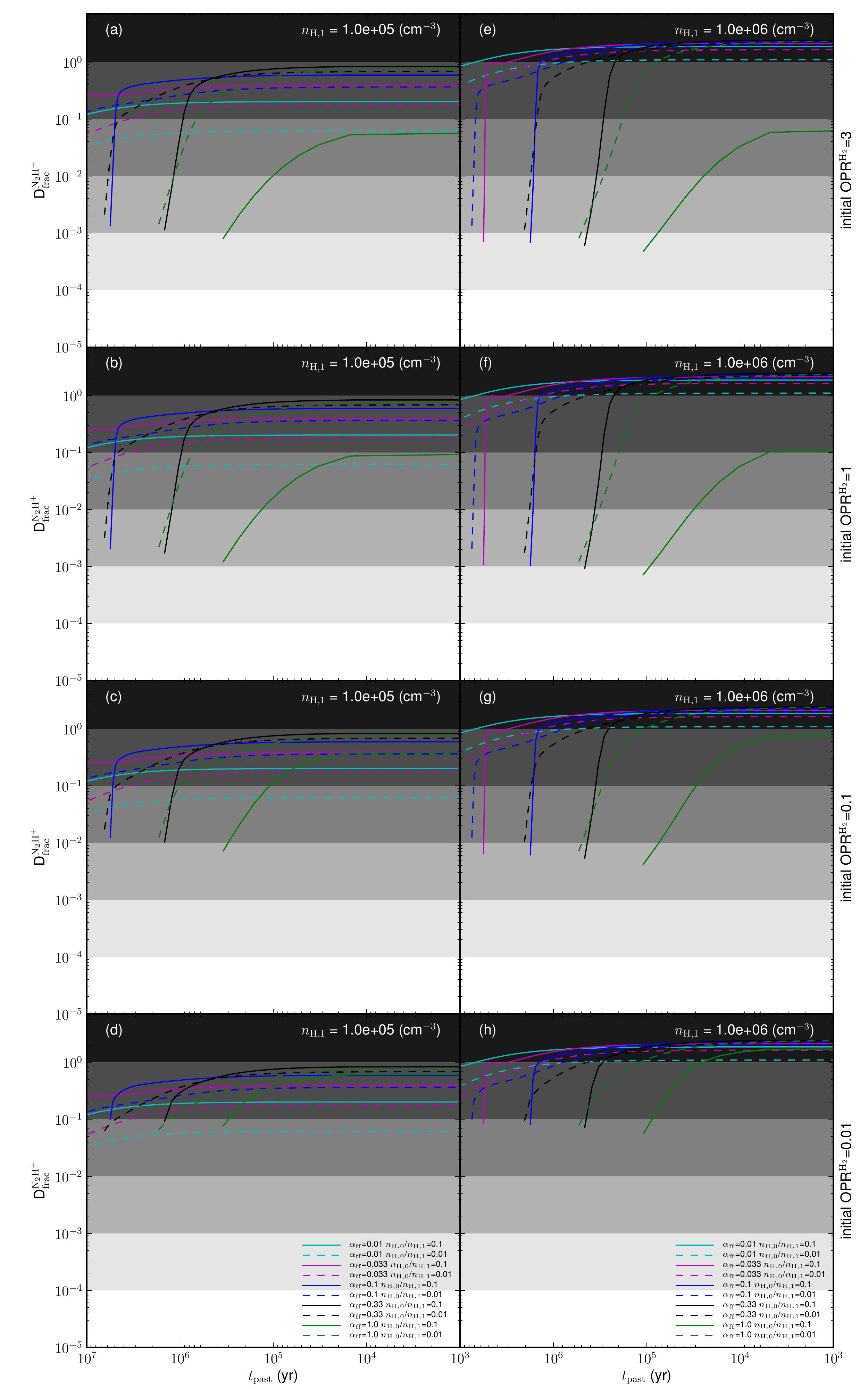}
\caption{
Same as Fig. \ref{fig:dynam_grain_op} but starting with $f_{\rm D}$=100.\label{fig:dynam_grain_op_fd100}}
\end{figure*}

We next explore the effect of the assumed starting OPR$_{\rm eq}^{\rm
  H_2}$ and the starting depletion factor. Focussing on models with
$n_{\rm H,1}=10^5$ and $10^6\:{\rm cm^{-3}}$ and with $\alpha_{\rm
  ff}=0.01, 0.033, 0.1, 0.33, 1$, we run TDD+DDE models for initial
OPR$_{\rm eq}^{\rm H_2}$=0.01, 0.1, 1, 3 and initial $f_{\rm D}$=1,10,100, and
show their results for $D_{\rm frac}^{\rm N_2H^+}$ in
Figures~\ref{fig:dynam_grain_op}, \ref{fig:dynam_grain_op_fd10},
\ref{fig:dynam_grain_op_fd100}.

For the models with initial $f_{\rm D}$=1, we find $D_{\rm frac}^{\rm
  N_2H^+}>0.1$ cores require $\alpha_{\rm ff}\lesssim 0.33$, 
  unless the starting OPR$_{\rm eq}^{\rm H_2}$=0.01.
However, these requirements become more relaxed if we start with
$f_{\rm D}$=10,100. To reconcile models of fast
collapse with high deuteration, it would require larger values of initial
$f_{\rm D}$ ($>$ 10) and small initial OPR$_{\rm eq}^{\rm H_2}$.

\begin{figure*}
\epsscale{1.2}
\plotone{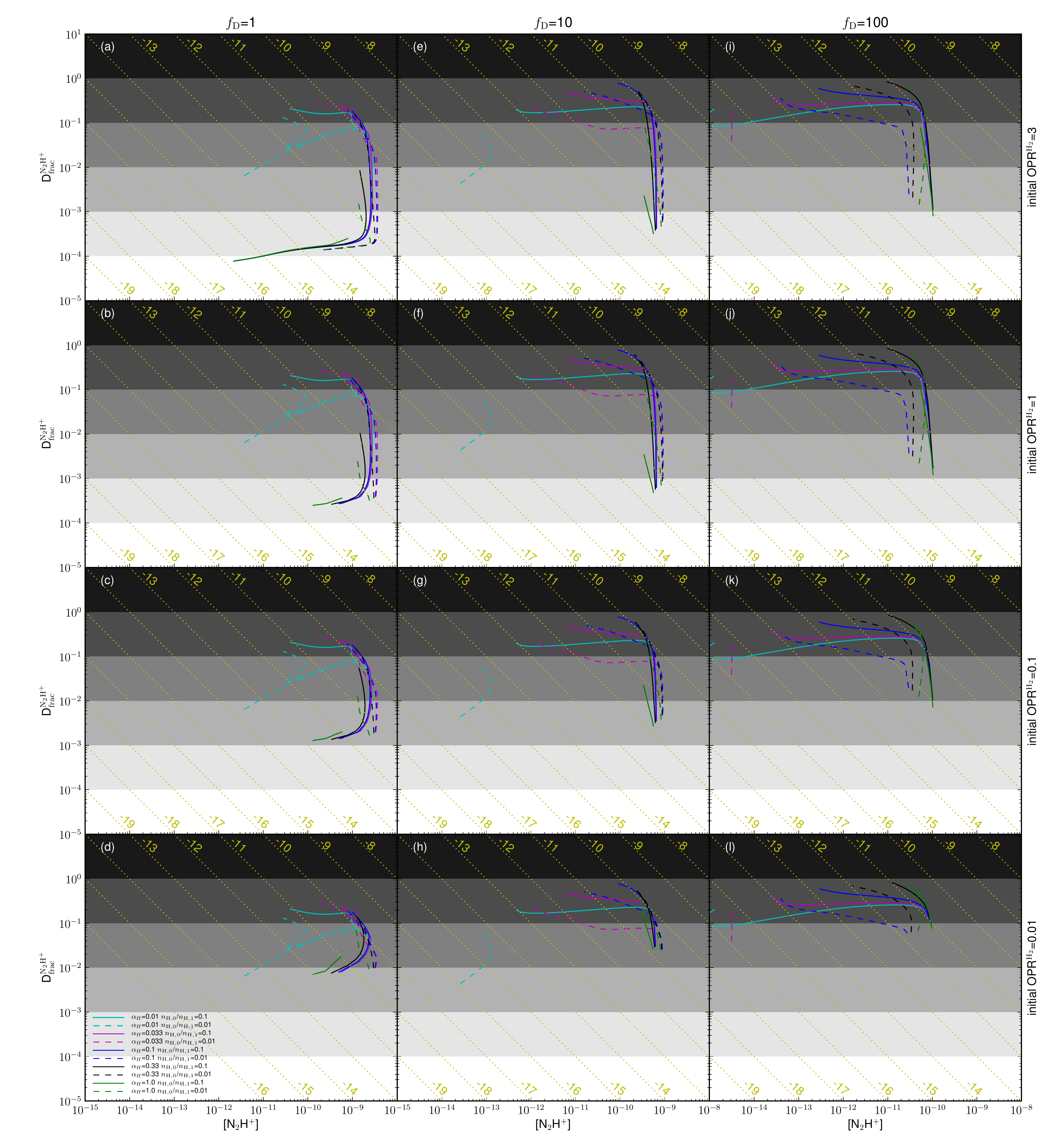}
\caption{
Relation between $D_{\rm frac}^{\rm N_2H^+}$ and [${\rm N_2H^+}$] with different starting $f_{\rm D}$ (labeled on top) and OPR$^{\rm H_2}$ (labeled on right). The modeled cores in this figure are from those in Figures~\ref{fig:dynam_grain_op},\ref{fig:dynam_grain_op_fd10},\ref{fig:dynam_grain_op_fd100}, having $n_{\rm H,1} = 10^5\:{\rm cm^{-3}}$. The yellow dotted lines show constant [${\rm N_2D^+}$] (the yellow numbers are indices with the base of 10).\label{fig:dynam_grain_op_fd_ntar1e5}}
\end{figure*}

\begin{figure*}
\epsscale{1.2}
\plotone{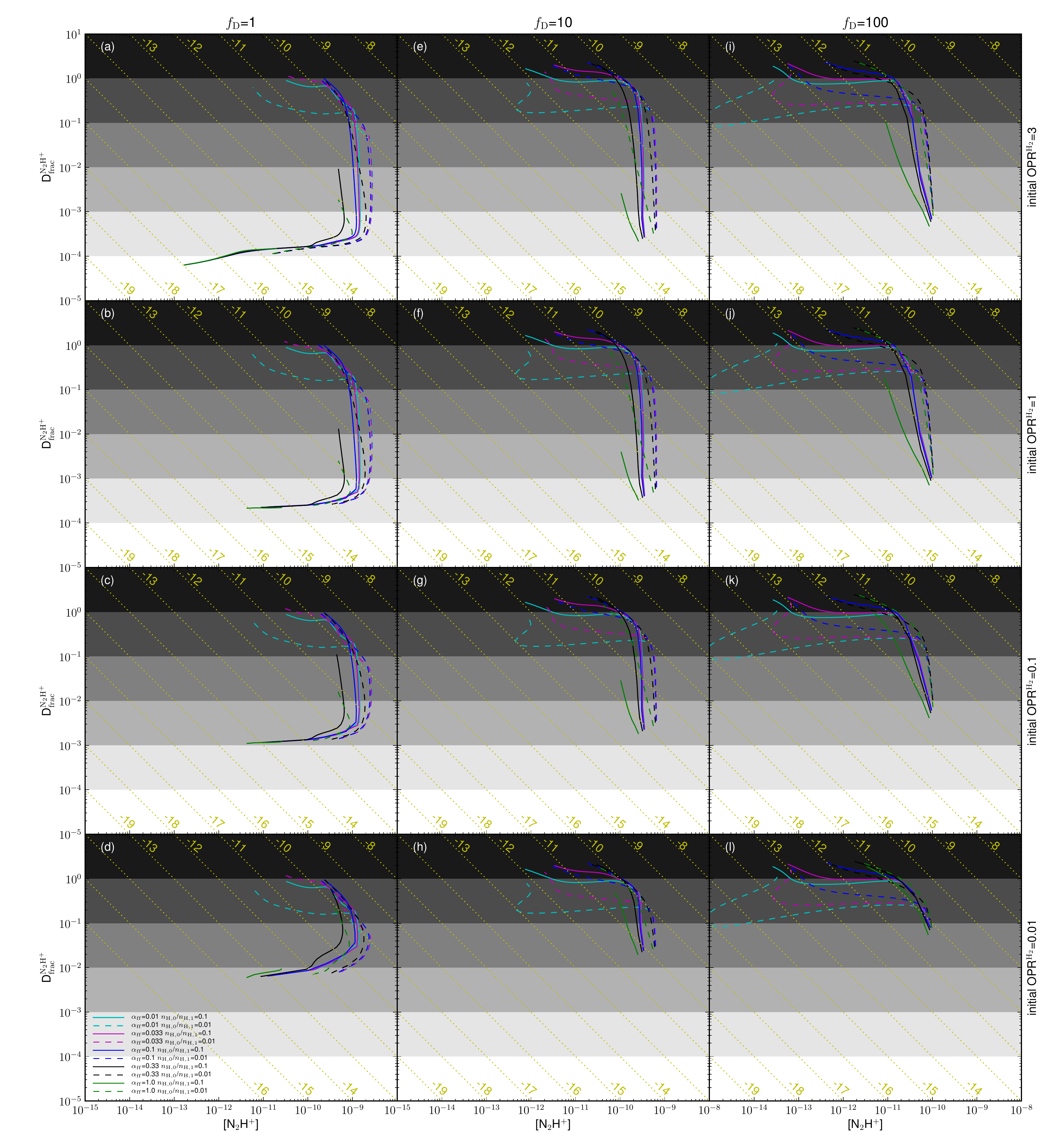}
\caption{
Same as Fig. \ref{fig:dynam_grain_op_fd_ntar1e5} but with $n_{\rm H,1} = 10^6\:{\rm cm^{-3}}$.\label{fig:dynam_grain_op_fd_ntar1e6}}
\end{figure*}

In order to see if these models with different starting values of
$f_{\rm D}$ can be separated by absolute abundances of [${\rm N_2H^+}$],
we plot the relation between $D_{\rm frac}^{\rm
  N_2H^+}$ and [${\rm N_2H^+}$] with $n_{\rm H,1} = 10^5\:{\rm
  cm^{-3}}$ in Figure~\ref{fig:dynam_grain_op_fd_ntar1e5} and $n_{\rm
  H,1} = 10^6\:{\rm cm^{-3}}$ in
Figure~\ref{fig:dynam_grain_op_fd_ntar1e6}. As expected, the absolute
abundances are lower in models with higher initial $f_{\rm D}$, so that
observations of these abundances, together with $D_{\rm frac}^{\rm
  N_2H^+}$, can help distinguish between the model families.

\section[]{Discussion} \label{sec:discussion}

\subsection{Deuteration as a Chemical Clock}

The high equilibrium values of $D_{\rm frac}^{\rm N_2H^+}$, together
with the small starting fractional abundance of D relative to H and
the controlling influence of the ortho-to-para ratio of $\rm H_2$,
which decays relatively slowly, mean that over a wide range of
parameter space relevant for cold, dense starless cores, the timescale
to reach deuteration equilibrium is relatively long compared to, for
example, the local free-fall timescale. Unless the starting conditions
for core formation involve extremely low values of OPR$^{\rm
  H_2}\lesssim 10^{-2}$, high values of depletion factor $f_{\rm D} \gtrsim
10$ or suffer high values of cosmic ray ionization $\zeta\gtrsim
10^{-16}\:{\rm s^{-1}}$, then observing high values of $D_{\rm
  frac}^{\rm N_2H^+}\gtrsim 0.1$ implies that the core is contracting
at rates significantly lower than free-fall, so that it has been in a dense,
cold state for at least several dynamical times. Note that if the core
is close to chemical equilibrium, then the derived deuteration
timescale is only a lower limit to its age. More accurate constraints
require tailored application of chemical models to particular physical
conditions of individual cores and may require measurement of absolute
abundances to constrain the effects of the depletion factor.

In one of the best studied low-mass pre-stellar cores, L1544 in the
Taurus molecular cloud, we can attempt to contrain an age.  Within the
central 3600\,AU (the beam size of the IRAM 30\,m antenna at the
frequency of the N$_2$H$^+$(1--0) line), $D_{\rm frac}^{\rm N_2H^+}$ =
0.2\footnote{We note that this observed $D_{\rm frac}^{\rm N_2H^+}$
  value should be treated as a lower limit, as this is an average
  along the line of sight and it is well known that the N$_2$H$^+$(1-0)
  emission is more extended than the N$_2$D$^+$(2-1) emission
  \citep{Casellietal2002}.  Therefore, our estimates of the time
  scales are also lower limits.}  \citep{Crapsi2005}, the average
number density is $n_{\rm H}\simeq$ 10$^6$\,cm$^{-3}$ \citep{2010MNRAS.402.1625K},
the temperature is about 6\,K \citep{Crapsi2007}, the cosmic ray
ionization rate is $\simeq$1$\times$10$^{-17}$\,s$^{-1}$ and the CO
depletion factor is 100 \citep{2010MNRAS.402.1625K}. With these parameters, we
obtain the time to reach $D_{\rm frac}^{\rm N_2H^+}$ =
0.2 to be 4.2$\times$10$^{5}$~yr --
2.6$\times$10$^{5}$~yr, starting with OPR$^{\rm H_2}$ = 3--0.1,
respectively, i.e. between 9.6 and 5.9 times the current value of
$t_{\rm ff}$.  The corresponding value of
OPR$^{\rm H_2}$ is expected to be $\sim 3\times$10$^{-3}$.

In cases where such a detailed analysis cannot be carried out, we can
still derive some limits on the core deuteration timescale. For
example, $D_{\rm frac}^{\rm N_2H^+} \gtrsim$ 0.1 has been measured in
low-mass pre-stellar cores (\citealt{Crapsi2005}; P09) and high-mass
starless cores \citep{Fontani2011,2012A&A...538A.137M}, and there are
currently no values of $D_{\rm frac}^{\rm N_2H^+}$ observed to be
greater than 1. 
Starting from this, we examine all our simple 
models used in our parameter space exploration (\S\ref{subsec:pe}) to
find how long it takes for $D_{\rm frac}^{\rm N_2H^+}$ to reach 0.1.
The results are shown as the blue dashed lines in the 4th row of
Fig. \ref{fig:exple}. The ``missing'' parts indicate conditions under
which $D_{\rm frac}^{\rm N_2H^+}$ fails to reach 0.1. As one can see
from the figure, to reach $D_{\rm frac}^{\rm N_2H^+} >$ 0.1, the cores
should be dense ($n_{\rm H} \gtrsim 5 \times$10$^4$ cm$^{-3}$), cold
($T \lesssim$ 17 K), at least moderately depleted ($f_{\rm D} \gtrsim$ 6),
and with moderate CR ionization rates ($\zeta \lesssim 6
\times$10$^{-17}$~s$^{-1}$). In all cases, the time to reach $D_{\rm
  frac}^{\rm N_2H^+}$ = 0.1 are longer than $t_{\rm ff}$. With
moderate depletion ($f_{\rm D} \lesssim$ 100), the large $D_{\rm
  frac}^{\rm N_2H^+}$ ($\gtrsim$ 0.1) is likely to indicate a large
deuteration age ($\gtrsim$ several $t_{\rm ff}$) for the observed
starless cores. As shown in Figure~\ref{fig:exple_op}, these constraints can be somewhat relaxed 
if the starting OPR$^{\rm H_2}$ values are small ($\lesssim 0.1$) and
the starting depletion factors large ($f_{\rm D} \gtrsim 100$) \citep[see
  also discussion in][and \S\ref{S:comparison}, below]{2011ApJ...739L..35P}.

\subsection[]{Implications for magnetic support and comparison with the ambipolar diffusion time}\label{subsec:tad}

If contraction of starless cores is very slow compared to the local
free-fall time, then this implies some form of pressure support is
resisting collapse \citep[see also][]{2010MNRAS.402.1625K}. 
In cores that are more massive than the thermal
Bonnor-Ebert mass, such as L1544 and the massive cores studied by Tan
et al. (2013), this pressure support would need to be nonthermal,
i.e. turbulence or magnetic fields. However, turbulence is expected to
decay relatively quickly, within $\sim 1 t_{\rm ff}$ 
\citep{1998ApJ...508L..99S,1998PhRvL..80.2754M}, leaving magnetic fields as
the favored option. This would imply core collapse occurs on the
ambipolar diffusion timescale $t_{\rm ad}$: the timescale for neutrals
in dense cores with low ionization fractions to contract relative to
the magnetic field \citep[e.g.][]{2004ApJ...616..283T}.

The ambipolar diffusion timescale can be calculated using the
expression $t_{\rm ad}$ = 2.5$\times$10$^{13} x(e)$ yr
\citep{spitzer1978,1987ARA&A..25...23S},
where $x(e)$ is the electron abundance relative to $n_{\rm H}$. Figure
\ref{fig:exple} plots $t_{\rm ad}$, to compare with $t_{\rm
  eq,90}$($D_{\rm frac}^{\rm N_2H^+}$) and $t_{\rm ff}$.  As $t_{\rm
  ad}$ is closely related to the ionization structure in core, high
density and low $\zeta$ conditions reduce $t_{\rm ad}$, as shown in
panels (e), (g), (m) and (o). For fiducial conditions, the deuteration
timescale is more similar to the local ambipolar diffusion
timescale than to the free-fall time.  Appreciating the caveats of
estimates of deuteration timescales, discussed above, we conclude this
is tentative, indirect evidence that magnetic fields are playing an
important role in regulating starless core formation, and thus star
formation.

\subsection{Comparison to Previous Studies}\label{S:comparison}

There have been a number of discussions regarding the 
deuteration chemistry in pre-stellar cores
(e.g. \citealt{1992A&A...258..479P}; \citealt{Flower2006}; P09; \citealt{2010A&A...509A..98S};
\citealt{2011A&A...526A..31P}; \citealt{2012ApJ...757L..11W}; \citealt{2013A&A...551A..38P}).
Compared with most of these previous studies, we have focussed on
$D_{\rm frac,eq}^{\rm N_2H^+}$, using a more complete cold core
chemistry network and with a larger and more systematic exploration of
the parameter space of environmental conditions that help control the
chemistry. 

In the following, we compare our model with some of these works.
\citet{2011A&A...526A..31P} benchmarked their results against that of P09 and
\citet{2010A&A...509A..98S}, but they do not show results for $D_{\rm
  frac}^{\rm N_2H^+}$.  Both P09 and \citet{2010A&A...509A..98S} used
\citet{Hugo2009} H$_3^+$ + H$_2$ reaction system (including spin
states and deuterium). \citet{2010A&A...509A..98S} also used dissociative
recombination reactions from P09, but no elements heavier than He were
considered. Taking this into consideration, we will only compare our
models directly with P09 out of these three papers. We will also
compare with \citet{2012ApJ...757L..11W}.
Note that although \citet{aikawa2012} built a comprehensive
chemical/dynamical model that included deuterium chemistry and
followed the evolution of pre-stellar cores to the formation of
protostars, since they do not include spin state chemistry, a direct
comparison with our results cannot be made.

Our work can be compared most closely to that of P09, 
who included spin state chemistry and modeled the evolution of
the abundance ratio of $\rm N_2D^+$ relative to $\rm N_2H^+$ and
discussed its use as a chemical clock. They used a modified version of
the Nahoon code to model about 35 species and 400 reactions. They did
not model N chemistry: in particular the abundance of $\rm N_2$ was a
parameter in their modeling, so absolute abundances are not predicted.
They developed a simple layered model for core structure that they
compared to observations of the pre-stellar core L183. Based on the
observed relatively low $D_{\rm frac}^{\rm N_2H^+}$ in the center of
the core, they concluded this central region must have only attained
high density relatively recently. They estimated a minimum age of
$\sim 2\times 10^5$~yr, but suggested that it may not be that much
older than this.

We ran our models with P09's choices of parameters, and compared with
their Figs. 7, 8 \& 9. In particular, they used $T$ = 7~K,
$\zeta$=2$\times$10$^{-17}$ s$^{-1}$. Their dust-to-gas mass ratio,
grain radius, and dust grain density are the same as our fiducial
model. They also used a fixed density and did not include
time-dependent depletion/desorption. They set CO abundance to be
10$^{-5}$ and $n_{\rm H}$=1.4$\times$10$^{5}$ cm$^{-3}$ in the outer
layer, and set CO abundance to be 10$^{-6}$ and $n_{\rm
  H}$=4.2$\times$10$^{6}$ cm$^{-3}$ in the inner layer. It is unclear
whether they had leftover C and O atoms in their models. Here we
simply assume all C and O were in CO in their models. Moreover, they
were unclear about what initial [N$_2$] they used in the models (this
was an input parameter of their models). We try our fiducial [N]
in both runs (case 1). 
We also tried two more models with [N$_2$] = [CO] value of P09 (case
2). We utilize P09 starting value of OPR$^{\rm H_2}$=3.

The results of our models are summarized in Table \ref{tab:comp} and
compared with P09. We first look at the timescales. For the outer
shell, the four timescales reported by our model are generally $\sim$
2-3 times that of P09 models. But for the inner shell our results are
closer to those of P09. Recalling our parameter space exploration in
Fig. \ref{fig:exple}, the depletion factor $f_{\rm D}$ has a strong
effect on the equilibrium timescales. Since it is unclear
what were the exact abundances used by P09 for C, N, O, the difference
of the timescales in Table \ref{tab:comp} could be due to differences
in abundances, i.e. depletion factor.

\begin{deluxetable*}{lcccccc}
\tabletypesize{\scriptsize}
\tablecaption{Comparison with P09 models of inner and outer shells.\label{tab:comp}}
\tablewidth{0pt}
\tablehead{
\colhead{Model}&\colhead{OPR$_{\rm eq}^{\rm H_2}$}&\colhead{$t_{\rm eq}$(OPR$^{\rm H_2}$)}&\colhead{$t_{\rm eq,90}$(OPR$^{\rm H_2}$)}&\colhead{$D_{\rm frac,eq}^{\rm N_2H^+}$}&\colhead{$t_{\rm eq}$($D_{\rm frac}^{\rm N_2H^+}$)}&\colhead{$t_{\rm eq,90}$($D_{\rm frac}^{\rm N_2H^+}$)}\\
\colhead{}&\colhead{($\times$10$^{-4}$)}&\colhead{(10$^6$ yr)}&\colhead{(10$^6$ yr)}&\colhead{}&\colhead{(10$^6$ yr)}&\colhead{(10$^6$ yr)}
}
\startdata
outer shell (P09)\tablenotemark{a} & 0.99 & 0.80 & 0.69 & 0.71 & 0.84 & 0.54 \\
outer shell (case1)\tablenotemark{b} & 1.15 & 1.90 & 1.28 & 0.399 & 1.64 & 1.02 \\
outer shell (case2)\tablenotemark{c} & 1.41 & 2.30 & 1.57 & 0.289 & 1.98 & 1.25 \\
\hline
inner shell (P09)\tablenotemark{a} & 0.53 & 0.38 & 0.28 & 5.9 & 0.30 & 0.17 \\
inner shell (case1)\tablenotemark{b}& 0.326 & 0.575 & 0.340 & 2.24 & 0.465 & 0.234 \\
inner shell (case2)\tablenotemark{c}& 0.299 & 0.515 & 0.298 & 2.96 & 0.415 & 0.200 
\enddata
\tablenotetext{a}{The P09 values are read from their figures using the Dexter tool incorporated in A\&A online journal, for which we estimate $\sim 1\%$ uncertainties.}
\tablenotetext{b}{In this case we used [N] in our fiducial model.}
\tablenotetext{c}{In this case we used [N] = [CO] in P09.}
\end{deluxetable*}

The fact that our fiducial model is somewhat slower compared to P09 Figure
7 is likely due to their choice of stronger depletion. 
Recalling our Fig. \ref{fig:exple}, the
high density in the P09 cores shortens the equilibrium timescale
somewhat, but this is compensated by P09's choice of a slightly
smaller $\zeta$. However, the large depletion factor can greatly
shorten the equilibrium time. In
P09 inner shell [CO]=10$^{-6}$ (if they did not have leftover atomic C
and O), then this corresponds to $f_{\rm D}$=146 for C and $f_{\rm
  D}$=360 for O based on our choices for initial elemental abundances,
which would greatly shorten the timescales if $f_{\rm D}$ for N is
comparable to our models.

In P09 Figure 7, the equilibrium time for $D_{\rm frac}^{\rm N_2H^+}$
was $\sim$ 5 times longer than the local instantaneous free-fall
time. However, their conclusion was that the core did not reach the
equilibrium. Our model predicts a much longer chemical equilibrium
timescale compared to the local free-fall time (also dependent on
$f_{\rm D}$). If we were to observe cores with the relevant high
$D_{\rm frac}^{\rm N_2H^+}$, then their ages should be relatively
old. 

Therefore, in terms of $D_{\rm frac,eq}^{\rm N_2H^+}$ and
$t_{\rm eq,90}$($D_{\rm frac}^{\rm N_2H^+}$), our model agrees
with P09 within a factor of 2-3. The major differences come from
the assumption of initial depletion and core equilibrium state.
These need to be constrained by observations.

\citet{2012ApJ...757L..11W} used a network with 4420 reactions, and their
equilibrium time for OPR$^{\rm H_2}$ is larger than 10$^6$ yr, which
is more similar to our value. They used $n_{\rm H}$=2$\times$10$^{6}$
cm$^{-3}$, $T$ = 10~K, $A_{\rm V}>$ 10~mag,
$\zeta$=3$\times$10$^{-17}$ s$^{-1}$, starting OPR$^{\rm H_2}$=3. They
allowed all neutral species (except for H$_2$, He, N, and N$_2$) to
freeze-out, but no desorption was considered. One thing to note is
that we are not sure about what initial elemental abundances
\citet{2012ApJ...757L..11W} used. They reference to Savage \& Sembach (1996)
who reported elemental abundances in a variety of environments. We are
not sure what specific initial abundances \citet{2012ApJ...757L..11W}
used. So the comparison here is just qualitative.

\section[]{Conclusions} \label{sec:conclusion}

We have presented a parameter space exploration of the deuterium
fractionation process, in particular of $\rm N_2H^+$, in conditions
appropriate to starless dense cloud cores in different environments.
An enhanced 3-atom reaction network is
introduced. It was derived from a reduced chemical network extracted
from the KIDA database to which Deuterium and spin state chemistry has
been included. Reactions involving H$_3$O$^+$ and its deuterated
forms are introduced from \citet{2013A&A...554A..92S}, to be able to reproduce 
results from the more comprehensive model of  \citet{2013A&A...554A..92S}.
The effects of time-dependent depletion and dynamical
density evolution have also been examined.
Compared to previous studies \citep[e.g.,][]{2009A&A...494..623P},
our focus is on conditions that are also relevant for massive star formation, 
such as the massive starless cores observed by \citet{2013ApJ...779...96T}.
Our main results are as follows:

\noindent
$\bullet$ 
Based on our fiducial modeling, the equilibrium value of [$\rm
  N_2D^+$]/[$\rm N_2H^+$] monotonically increases with increasing
density (from 10$^3$ cm$^{-3}$ $<$ $n_{\rm H}$ $<$ 10$^7$ cm$^{-3}$),
and decreasing CR ionization rate (10$^{-18}$ s$^{-1}$ $<$
$\zeta$ $<$ 10$^{-15}$ s$^{-1}$).  With increasing temperature, the
equilibrium [$\rm N_2D^+$]/[$\rm N_2H^+$] first moderately increases from $T$
$\simeq$ 5~K to $T$ $\simeq$ 13~K, then decreases to $T \simeq$ 30~K.
With increasing freeze-out, the equilibrium [$\rm N_2D^+$]/[$\rm N_2H^+$] 
increases from $f_{\rm D}$ $\simeq$ 1 to $f_{\rm D}$ $\simeq$ 1000,
but drops from $f_{\rm D}$ $\simeq$ 2000 to $f_{\rm D}$ $\simeq$ 10$^6$.

\noindent
$\bullet$ When the gas temperature exceeds $\simeq$ 20\,K, the
ortho-to-para H$_2$ ratio increases, reducing the deuterium fraction,
so that warmer starless cores should display lower deuterium fractions
(as found in high-mass star-forming regions by Fontani et al. 2011).

\noindent
$\bullet$ The above findings are robust against changes in the initial
elemental and molecular abundances.

\noindent
$\bullet$ Constraints on core ages and collapse rates can be obtained
if accurate measurements of [$\rm N_2D^+$]/[$\rm N_2H^+$] are made,
coupled with observations of core density, temperature and (CO)
depletion structure. However, results can also depend on the
cosmic ray ionization rate and the initial ortho-to-para ratio of $\rm
H_2$. 

\noindent
$\bullet$ In the case of the well-known low-mass pre-stellar core
L1544, we estimate that the gas within the central 3600\,AU has a
deuteration age between $\simeq$6 and 10 times the current local
free-fall time, depending on the initial value of the ortho-to-para
H$_2$ ratio.

\noindent
$\bullet$ More generally, to reproduce the typical deuterium fractions
measured toward low-mass and massive pre-stellar cores ([$\rm
  N_2D^+$]/[$\rm N_2H^+$]$\ga$ 0.1), the following physical parameters
are needed: $n_{\rm H}$ $\ga$ 3$\times$10$^4$ cm$^{-3}$, $T$ $\la$ 17
K, depletion factor $\ga$ 6, and cosmic ray ionization rate $\la$
10$^{-16}$ s$^{-1}$. In general, these values of deuterium
fractions require timescales several times longer than the local
free-fall timescale.  
With no initial depletion, the inclusion of time-dependent
depletion/desorption has only a modest effect on these conclusions. 
Also with no initial depletion, models with
dynamically evolving density, increasing by a factor of 10, require collapse rates about 10 times
slower than free-fall to reach the above levels of deuteration in
cores with $n_{\rm H} = 10^6\:{\rm cm^{-3}}$. This suggests that
dense cores with large deuterium fractions are dynamically old, which
would likely require support against gravity to be provided by
magnetic fields. For our fiducial model parameters, the timescale to
reach deuteration equilibrium is similar to the expected ambipolar
diffusion timescale, i.e., the collapse time of a magnetically
subcritical core. The above conclusions can be avoided if 
the initial depletion factor is $\ga$ 10 (in which case rapidly collapsing
cores could reach [$\rm N_2D^+$]/[$\rm N_2H^+$]$\ga$ 0.1), the cosmic
ray ionization rate is very high ($\gtrsim 10^{-16}\:{\rm s}^{-1}$)
or if the initial ortho-to-para ratio of $\rm H_2$ in the core is very
small ($\lesssim 0.01$), although this last condition itself would require
the parental cloud to have a significant age.

\section*{Acknowledgments}
The authors acknowledge the continuous and fruitful interactions with
Jorma Harju, and an anonymous referee for helping improve the manuscript. 
SK acknowledges support from Xueying Tang
and an NRAO Student Observing Support grant. JCT acknowledges support
from Univ. of Florida Research Opportunity Seed Fund and the Florida
Space Inst. VW acknowldeges funding by the French INSU/CNRS program
PCMI, the Observatoire Aquitain des Sciences de l'Univers and the
European Research Council (ERC Grant 336474: 3DICE).

\newpage

\label{lastpage}

\end{document}